\pdfoutput=1

\documentclass[11pt]{article}
\usepackage{cite}
\usepackage{graphicx}
\usepackage{amsmath}
\usepackage{amsfonts}
\usepackage{amssymb}
\usepackage[usenames,dvipsnames]{color}

\newcommand{\be}{\begin{equation}}
\newcommand{\ee}{\end{equation}}
\newcommand{\ba}{\begin{eqnarray}}
\newcommand{\ea}{\end{eqnarray}}
\newcommand{\nn}{\nonumber}
\newcommand{\barr}{\begin{array}}
\newcommand{\earr}{\end{array}}

\newcommand\lsim{\mathrel{\rlap{\lower4pt\hbox{\hskip1pt$\sim$}}
        \raise1pt\hbox{$<$}}}
\newcommand\gsim{\mathrel{\rlap{\lower4pt\hbox{\hskip1pt$\sim$}}
        \raise1pt\hbox{$>$}}}

\def\fnl{f_{NL}}
\def\gnl{g_{NL}}

\def\x{{\bf x}}

\def\k{{\bf k}}

\def\q{{\bf q}}
\def\r{{\bf r}}
\def\bigoh{{\mathcal O}}
\def\erfc{\mbox{erfc}}

\begin{document}

\begin{titlepage}
\setcounter{page}{1} \baselineskip=15.5pt \thispagestyle{empty}

\bigskip\
\begin{center}
{\Large \bf Halo clustering and $\gnl$-type primordial non-Gaussianity}
\end{center}
\vspace{0.5cm}
\begin{center}
{\fontsize{14}{30}\selectfont
  Kendrick M.~Smith$^1$,
  Simone Ferraro$^1$, and
  Marilena LoVerde$^2$
}
\end{center}

\begin{center}
\vskip 8pt
\textsl{${}^1$ Princeton University Observatory, Peyton Hall, Ivy Lane, Princeton, NJ 08544 USA}
\vskip 4pt
\textsl{${}^2$ Institute for Advanced Study, Einstein Drive, Princeton, NJ 08540, USA}
\end{center}
\vspace{1.2cm}

\hrule \vspace{0.3cm}
{ \noindent \textbf{Abstract} \\[0.2cm]
\noindent
A wide range of multifield inflationary models generate non-Gaussian initial conditions
in which the initial adiabatic fluctuation is of the form $(\zeta_G + \gnl \zeta_G^3)$.
We study halo clustering in these models using two different analytic methods: the peak-background
split framework, and brute force calculation in a barrier crossing model, obtaining agreement
between the two.
We find a simple, theoretically motivated expression for halo bias which agrees
with $N$-body simulations and can be used to constrain $\gnl$ from observations. 
We discuss practical caveats to constraining $\gnl$ using only observable properties of a
tracer population, and argue that constraints obtained from populations whose observed bias
is $\lsim 2.5$ are generally not robust to uncertainties in modeling the halo occupation
distribution of the population.}
 \vspace{0.3cm}
 \hrule

\vspace{0.6cm}
\end{titlepage}
\newpage

\section{Introduction}
\label{sec:intro}

In the last few decades, increasingly precise observations
(e.g.~\cite{Komatsu:2010fb,Percival:2009xn,Reid:2009xm,Riess:2011yx,Kessler:2009ys,Vikhlinin:2008ym})
have led to a standard cosmological model in which
small initial fluctuations evolve in a $\Lambda$CDM background to 
give rise to the observed universe.
Current data are consistent with initial fluctuations which are adiabatic, scalar, Gaussian, with weak
deviations from scale invariance ($n_s < 1$ at 3$\sigma$).

The statistics of the initial fluctuations, i.e. deviations from Gaussian initial conditions,
 provide a powerful probe of the physics of the early universe.
In the context of inflation 
\cite{Guth:1980zm,Linde:1983gd,Albrecht:1982wi,Guth:1982ec,Hawking:1982cz,Starobinsky:1982ee,Bardeen:1983qw},
the simplest models (single-field, minimally coupled slow-roll) predict initial curvature perturbations
 with negligible deviations from Gaussianity.  
However, there is a rich phenomenology of non-Gaussian initial conditions in
models with multiple fields, self-interactions near horizon crossing, or speed of sound $c_s \ll 1$ during inflation.
In this paper, we will focus on non-Gaussianity of the so-called local type \cite{Salopek:1990jq,Gangui:1993tt,Komatsu:2001rj,Okamoto:2002ik},
in which the primordial potential\footnote{In studies of primordial non-Gaussianity, 
it is conventional to define a primordial potential $\Phi = \frac{3}{5}\zeta$, where $\zeta$ is the initial adiabatic curvature 
perturbation.  Note that $\Phi$ is not the conformal Newtonian potential, which is given by $\frac{2}{3}\Phi$ deep in the
radiation-dominated epoch where Eq.~(\ref{eq:local_ng}) applies.}
is of the form
\be
\Phi(\x) = \Phi_G(\x) + \fnl (\Phi_G(\x)^2 - \langle \Phi_G^2 \rangle )
  + \gnl (\Phi_G(\x)^3 - 3 \langle \Phi_G^2 \rangle \Phi_G(\x) )  \label{eq:local_ng}
\ee
where $\Phi_G$ is a Gaussian field and $\fnl$, $\gnl$ are free parameters.\footnote{We define
$\gnl$-type non-Gaussianity including the term $-3\langle\Phi_G^2\rangle\Phi_G$;
this term simply renormalizes $\Phi_G$ so that its power spectrum $P_{\Phi_G}$ is equal
to the observed power spectrum $P_\Phi$ (to first order in $\gnl$).}

Local non-Gaussianity can be generated by physical mechanisms involving multiple fields, such as
light spectator fields during inflation which evolve to generate the initial adiabatic fluctuations (the 
curvaton scenario) \cite{Linde:1996gt,Lyth:2001nq,Lyth:2002my}, or models where the inflaton decay rate
is modulated by a second field \cite{Dvali:2003em,Kofman:2003nx}.
Non-Gaussianity of local type is also naturally generated in non-inflationary models of the early universe
such as the new ekpyrotic/cyclic scenario \cite{Buchbinder:2007at,Creminelli:2007aq,Lehners:2007wc}.
There is a theorem \cite{Maldacena:2002vr,Creminelli:2004yq}
which states that any single-field model of inflation cannot generate detectable levels of local
non-Gaussianity without violating observed limits on deviation from a scale-invariant power spectrum.
Thus, detection of either $\fnl$ or $\gnl$ would rule out all single field models of inflation and
place powerful constraints on the physics of the early universe.
Current observational constraints on these parameters are consistent with zero 
\cite{Komatsu:2010fb,Slosar:2008hx,Fergusson:2010gn,Desjacques:2008vf}, but are expected
to improve by an order of magnitude or more in the near future.

In models of inflation in which $|\gnl| = \bigoh(\fnl^2)$, it is unlikely that observational
constraints on $\gnl$ will be competitive with constraints on $\fnl$.
However, there are a number of examples where $\fnl^2 \ll|\gnl| $, 
 making the $\gnl$ term in Eq.~(\ref{eq:local_ng})
the dominant source of primordial non-Gaussianity. This situation arises in 
curvaton models where non-quadratic terms in the potential are important 
\cite{Sasaki:2006kq,Ichikawa:2008iq,Enqvist:2008gk,Huang:2008zj,Chingangbam:2009xi}
or in multifield models in which $(\Delta N)$ varies rapidly at the end of inflation
\cite{Huang:2009vk,Byrnes:2009qy}. The existence of these scenarios makes searching for
$\gnl$ just as important as $\fnl$ and measurements
provide important constraints on the microphysical parameter space.

In a pioneering paper \cite{Dalal:2007cu}, Dalal et al showed that large-scale clustering
of halos depends sensitively on $\fnl$.  More precisely, a sample of halos (or tracers such as
galaxies or quasars) with constant bias $b_1$ in a Gaussian cosmology will have scale-dependent
bias given by
\be
b(k) \approx b_1 + 2\delta_c (b_1-1) \frac{\fnl}{\alpha(k,z)}   \label{eq:ddhs}
\ee
in an $\fnl$ cosmology.  Here, $\delta_c$ is the spherical collapse threshold and
$\alpha(k,z)$ is defined by
\be
\alpha(k,z) = \frac{2 k^2 T(k) D(z)}{3 \Omega_m H_0^2}  \label{eq:alpha_def}
\ee
so that the linear density field and the primordial potential are related by
$\delta_{\rm lin}(k,z) = \alpha(k,z) \Phi(k)$.
Large-scale structure constraints on $\fnl$ from scale-dependent bias are currently
competitive with the CMB (e.g.~\cite{Slosar:2008hx,Xia:2011hj}) and may ultimately provide
constraints which are stronger (e.g.~\cite{Cunha:2010zz,Hamaus:2011dq}).
The key identity~(\ref{eq:ddhs}) has been derived using several different analytic
frameworks \cite{Matarrese:2008nc,Slosar:2008hx,Giannantonio:2009ak}
 and agrees with $N$-body simulations (e.g. 
 \cite{Dalal:2007cu,Grossi:2007ry,Pillepich:2008ka,Desjacques:2008vf}).

In this paper we study the related issue of large-scale halo-clustering in a $\gnl$
cosmology.  We consider the large-scale halo bias in two analytic frameworks:
 the peak-background split 
 (\S\ref{sec:pbs}) and a barrier crossing model (\S\ref{sec:barrier}).  We find consistency
 between the two formalisms (in disagreement with \cite{Desjacques:2011mq})
and obtain an expression analogous to Eq.~(\ref{eq:ddhs}) for the 
scale-dependent halo bias in a $\gnl$ cosmology. Our main results are a universal relation 
between the scale-dependent halo bias in a $\gnl$ cosmology and the mass function in an $\fnl$ cosmology, 
\be
b(k)\approx b_1+\frac{\beta_{g}\gnl}{\alpha(k,z)}\,\quad \textrm{where}\quad \, \beta_{g}=3(\partial\log n/\partial\fnl)
\label{eq:bgnl}
\ee
and expressions for $\beta_g$ (Eqs.~(\ref{eq:b2_final}),~(\ref{eq:b2_mw}))
which can be used in practice to constrain $\gnl$ from data.  
We also discuss caveats when estimating the $\gnl$ bias from observable
quantities (\S\ref{ssec:caveat}) and argue that constraints obtained from
tracer populations which are not highly biased ($b_1 \gsim 2.5$) are generally
not robust to uncertainties in HOD modeling.

Throughout this paper we use the WMAP5+BAO+SN fiducial cosmology \cite{Dunkley:2008ie}, with
baryon density $\Omega_bh^2 = 0.0226$, CDM density $\Omega_ch^2 = 0.114$, Hubble parameter $h=0.70$,
spectral index $n_s=0.961$, optical depth $\tau = 0.080$, and power-law initial curvature power spectrum 
$k^3 P_\zeta(k) / 2\pi^2 = \Delta_\zeta^2 (k/k_{\rm piv})^{n_s-1}$ where $\Delta_\zeta^2 = 2.42 \times 10^{-9}$
and $k_{\rm piv} = 0.002$ Mpc$^{-1}$.
All power spectra and transfer functions have been computed using CAMB \cite{Lewis:1999bs}.

\section{Definitions and notation}
\label{sec:notation}

We will sometimes model halos of mass $\ge M$ with peaks in a smoothed density field $\delta_M$ defined as follows.
Let $\delta_M(\x)$ be the {\em linear} density field smoothed by a tophat filter with radius $R(M) = (3M/4\pi\rho_m)^{1/3}$, i.e.
\be
\delta_M(\x) = \int \frac{d^3\k}{(2\pi)^3} e^{-i\k\cdot\x} \delta_{\rm lin}(\k) W_M(k)   \label{eq:deltaM_def}
\ee
where
\be
W_M(k) = 3\, \frac{\sin(kR(M)) - kR(M) \cos(kR(M))}{(kR(M))^3}
\ee
Let $\sigma_M = \langle \delta_M^2 \rangle^{1/2}$  be the RMS amplitude of the smoothed density field, and
let $\kappa_n(M)$ be its $n$-th non-Gaussian cumulant, defined by:
\be
\kappa_n(M) = \frac{\langle \delta_M^n \rangle_{\rm conn}}{\sigma_M^n}  \label{eq:kappa_def}\,.
\ee
Since $\delta_M$ and $\sigma_M$ are defined via linear theory, $\kappa_n(M)$ is independent of redshift
as implied by the notation.
To first order in $\fnl$ and $\gnl$, we have
\ba
\kappa_3(M) &=& \kappa_3^{(1)}(M) \fnl \\
\kappa_4(M) &=& \kappa_4^{(1)}(M) \gnl
\ea
with higher cumulants equal to zero,
where $\kappa_3^{(1)}(M), \kappa_4^{(1)}(M)$ are the values of the cumulants at $\fnl=1$ and $\gnl=1$ respectively.
These values are given explicitly by:
\ba
\kappa_3^{(1)}(M) &=& \frac{6}{\sigma_M^3} \int \frac{d^3\k\,d^3\k'}{(2\pi)^6} W_M(k) W_M(k') W_M(|\k+\k'|) 
    \frac{P_{mm}(k) P_{mm}(k') \alpha(|\k+\k'|)}{\alpha(k) \alpha(k')} \label{eq:kappa3_explicit} \\
\kappa_4^{(1)}(M) &=& \frac{24}{\sigma_M^4} \int \frac{d^3\k\,d^3\k'\,d^3\k''}{(2\pi)^9} 
   W_M(k) W_M(k') W_M(k'') W_M(|\k+\k'+\k''|) \nn \\
  && \hspace{2cm} \times \frac{P_{mm}(k) P_{mm}(k') P_{mm}(k'') \alpha(|\k+\k'+\k''|)}{\alpha(k) \alpha(k') \alpha(k'')}
\ea
where $\alpha(k)$ was defined previously in Eq.~(\ref{eq:alpha_def}) and $P_{mm}(k)$ is the power spectrum
 of the linear density field, $\langle \delta_{\rm lin}(\k) \delta_{\rm lin}(\k')\rangle=(2\pi)^3P_{mm}(k)\delta^{(3)}(\k+\k')$.
For numerical calculation, the following fitting functions (from \cite{LoVerde:2011iz}) are convenient:
\ba
\kappa_3^{(1)}(M) &=& (6.6\times 10^{-4}) \left( 1 - 0.016 \log\left(\frac{M}{h^{-1} M_\odot}\right) \right) \\
\kappa_4^{(1)}(M) &=& (1.6 \times 10^{-7}) \left(1 - 0.021 \log \left(\frac{M}{h^{-1}M_\odot}\right) \right)\,.
\ea
This paper is mainly concerned with calculating halo bias $b(k) = P_{mh}(k)/P_{mm}(k)$ to first order in
$\fnl$ and $\gnl$, so let us establish notation from the outset, by writing the large-scale bias in the general form:
\be
b(k) = b_1 + b_{1f} \fnl + b_{1g} \gnl + \frac{\beta_f \fnl + \beta_g \gnl}{\alpha(k)}
\label{eq:bdefs}
\ee
where unlike Eq.~(\ref{eq:ddhs}) and Eq.~(\ref{eq:bgnl}) we have allowed for
 scale-independent corrections $b_{1f}$ and $b_{1g}$ 
 from $\fnl$ and $\gnl$ primordial non-Gaussianity. Equation~(\ref{eq:bdefs}) 
 defines the coefficients $b_1, b_{1f}, b_{1g}, \beta_f, \beta_g$.
This equation assumes that the $k$-dependence is of the functional form 
$(\mbox{constant}) + (\mbox{constant})/\alpha(k)$,
but we will derive this analytically (Eq.~(\ref{eq:barrier_bias})) and show that
it agrees with simulations (\S\ref{ssec:bias_fits}).
In this notation, the Dalal et al formula~(\ref{eq:ddhs}) can be written as $\beta_f = 2\delta_c(b_1-1)$.

\section{Peak-background split}
\label{sec:pbs}

The peak-background split formalism is a procedure for predicting halo clustering statistics on large scales.
The basic idea is that a long-wavelength fluctuation in the initial curvature alters the local abundance of halos in a
way which is equivalent to perturbing parameters of the background cosmology, e.g.~the matter density $\rho_m$
or the amplitude $\Delta_\Phi$ of the initial fluctuations.
The use of this formalism to study halo bias in non-Gaussian cosmologies was pioneered in \cite{Slosar:2008hx};
we will review this calculation of the bias in an $\fnl$ cosmology (\S\ref{ssec:pbs_fnl}) and then perform an analogous
calculation in the $\gnl$ case (\S\ref{ssec:pbs_gnl}).

\subsection{$\fnl$ cosmology}
\label{ssec:pbs_fnl}

In an $\fnl$ cosmology, the initial conditions are given by:
\be
\Phi(\x) = \Phi_G(\x) + \fnl(\Phi_G(\x)^2 - \langle \Phi_G^2 \rangle)  \label{eq:pbs_fnl}
\ee
To analyze the effect of a long-wavelength mode, let us 
decompose the {\em Gaussian} potential as a sum $\Phi_G = \Phi_l + \Phi_s$ of long-wavelength and short-wavelength 
contributions. The long/short-wavelength decomposition of the non-Gaussian potential $\Phi$ is then
\be
\Phi(\x) = \underbrace{ \Phi_l(\x) + \fnl \left(\Phi_l(\x)^2 - \langle \Phi_l^2 \rangle \right) }_{\rm long}
  + \underbrace{(1 + 2\fnl\Phi_l(\x)) \Phi_s(\x) + \fnl( \Phi_s(\x)^2 - \langle \Phi_s^2 \rangle )}_{\rm short}
  \label{eq:longshort}
\ee
and contains explicit coupling between long and short wavelength modes of the Gaussian potential. 

Let us consider how the term $(1 + 2\fnl\Phi_l(\x)) \Phi_s(\x)$ in Eq.~(\ref{eq:longshort})
affects $n_l(\x)$, the long-wavelength part of the halo number density field.
In a local region where the long-wavelength potential takes some value $\Phi_l$, the amplitude $\Delta_\Phi$ of the small-scale
modes is perturbed: $\Delta_\Phi\rightarrow (1+2\fnl\Phi_l)\Delta_\Phi$.
This modifies the local halo abundance, in the same way that the global abundance would be modified if the cosmological parameter
$\Delta_\Phi$ were perturbed, i.e.~we get a term in the long-wavelength halo density of the form
$\Delta n(\x) = 2 \fnl \Phi_l(\x) (\partial n/\partial \log \Delta_\Phi)$.
In addition, even in a Gaussian cosmology, there is a perturbation to the local halo abundance which is proportional to the
long-wavelength part $\delta_l(\x)$ of the density fluctuation, i.e.~a term of the form
$\Delta n(\x) = \delta_l(\x) (\partial n/\partial\delta_l)$.
Putting this together, the long-wavelength part of the halo density is given by:\footnote{In this derivation,
we have swept two terms in Eq.~(\ref{eq:pbs_fnl}) under the rug; let us now argue that these are indeed negligible.
The term $\fnl (\Phi_s(\x)^2 - \langle \Phi_s^2\rangle)$ alters the statistics of the small scale modes;
this does perturb the halo abundance (by generating skewness in the density field)
but the perturbation is independent of the long-wavelength fluctuation $\Phi_l$.
Therefore, this term does not contibute to the large-scale halo bias.
The term $\fnl (\Phi_l(\x)^2 - \langle \Phi_l^2 \rangle)$ perturbs the long-wavelength modes and
decorrelates them (to order $\bigoh(\fnl)$) from both the linear density fluctuation $\delta(\x)$
and the field $(2 \fnl \Phi_l)$ which modulates the local power spectrum
amplitude $\Delta_\Phi$.
In principle, this should generate stochastic bias at order $\bigoh(\fnl^2)$, but we will neglect this, since we are
only calculating to order $\bigoh(\fnl)$.}
\ba
n_l(\x) &=& {\bar n} + \frac{\partial n}{\partial \delta_l} \delta_l(\x) + 2 \fnl \frac{\partial n}{\partial \log \Delta_\Phi} \Phi_l(\x) \nn \\
  &=& {\bar n} ( 1 + b_1 \delta_l(\x) + \beta_f \fnl \Phi_l(\x))  \label{eq:pbs_fnl_nl}
\ea
where
\ba
b_1 &=& \frac{\partial\log n}{\partial\delta_l}  \label{eq:pbs_b0def} \\
\beta_f &=& 2 \frac{\partial\log n}{\partial \log \Delta_\Phi} \label{eq:pbs_b1def}\,.
\ea
Intuitively, in an $\fnl$ cosmology, the local power spectrum amplitude $\Delta_\Phi$ is not spatially constant, but varies
throughout the universe in a way which is proportional to the long-wavelength potential $\Phi_l$.

Computing the halo bias $b(k) = P_{mh}(k)/P_{mm}(k)$ from Eq.~(\ref{eq:pbs_fnl_nl}) for $n_l(\x)$, we get:
\ba
b(k) 
  &=& \frac{b_1 P_{mm}(k) + \beta_f P_{m\Phi}(k)}{P_{mm(k)}} \nn \\
  &=& b_1 + \frac{\beta_f \fnl}{\alpha(k,z)}  \label{eq:pbs_fnl_bweak}\,.
\ea
From the preceding argument, we predict that the scale-dependent $\fnl$ bias is given by
$\beta_f = 2 (\partial\log n/\partial\log\Delta_\Phi)$.  We will refer to this as a ``weak''
prediction for the bias: it cannot be used to constrain $\fnl$ from real data, since $\beta_f$
has not been expressed in terms of observable quantities.

To make further progress, we need to evaluate the derivative $(\partial\log n/\partial\log\Delta_\Phi)$,
by making additional assumptions.
If we assume that the halo mass function is universal,
then one can calculate the derivative, obtaining 
$(\partial\log n/\partial\log\Delta_\Phi) = \delta_c (b_1-1)$, where
$b_1$ is the Gaussian bias \cite{Slosar:2008hx}, 
so that:
\be
\beta_f = 2\delta_c(b_1-1)\,.
\ee
We will refer to this as a ``strong'' prediction for the scale-dependent bias in an $\fnl$ cosmology,
since $\beta_f$ has been expressed in terms of the observable quantity $b_1$.  The strong form is essential
for constraining $\fnl$ from observations.

\subsection{$\gnl$ cosmology}
\label{ssec:pbs_gnl}

Let us now generalize the analysis of large-scale clustering in the previous subsection to the case
of a $\gnl$ cosmology, with initial conditions given by:
\be
\Phi(\x) = \Phi_G(\x) + \gnl(\Phi_G(\x)^3 - 3 \langle \Phi_G^2 \rangle \Phi_G(\x))\,.
\ee
Separating the Gaussian field into long and short wavelength pieces $\Phi_G=\Phi_l+\Phi_s$,
we decompose $\Phi$ as follows:
\ba
\label{eq:pbs_gnl}
\Phi(\x) &=& \underbrace{ \Phi_l(\x) + \gnl(\Phi_l(\x)^3 - 3\langle\Phi_l^2\rangle \Phi_l(\x))}_{\rm long}    \\
   &+& \underbrace{\Phi_s(\x) 
              + 3 \gnl(\Phi_l(\x)^2 - \langle\Phi_l^2\rangle) \Phi_s(\x) 
              + 3 \gnl \Phi_l(\x) (\Phi_s(\x)^2 - \langle\Phi_s^2\rangle)
              + \gnl (\Phi_s(\x)^3 - 3 \langle \Phi_s^2 \rangle \Phi_s(\x))\nn
     }_{\rm short}
\ea
As in the $\fnl$ case, we'll consider the perturbation to the long-wavelength halo density $n_h(\x)$ generated by each of these terms.

The term $3 \gnl(\Phi_l(\x)^2 - \langle\Phi_l^2\rangle) \Phi_s(\x)$ can be interpreted as a local modulation in the
small-scale power spectrum amplitude, given by $\Delta_\Phi \rightarrow (1 + 3\gnl(\Phi_l(\x)^2 - \langle\Phi_l^2\rangle)) \Delta_\Phi$.
This generates a term $\Delta n_l(\x) = 3\gnl(\Phi_l(\x)^2 - \langle\Phi_l^2\rangle) (\partial n/\partial\log\Delta_\Phi)$ in the
long-wavelength halo density, in close analogy with the $\fnl$ case (the modulation is proportional
to $\gnl (\Phi_l^2 - \langle\Phi_l^2\rangle)$ in this case, rather than $\fnl \Phi_l$).

The term $3 \gnl \Phi_l(\x) (\Phi_s(\x)^2 - \langle\Phi_s^2\rangle)$ can be interpreted as follows.
In a local region where the long-wavelength potential takes the value $\Phi_l$, the small-scale modes are perturbed in the
same way as in an $\fnl$ cosmology where the global value of $\fnl$ is given by $(3 \gnl \Phi_l)$.
This generates a term $\Delta n_l(\x) = 3 \gnl \Phi_l(\x) (\partial n/\partial\fnl)$ in the long-wavelength halo density.

Finally, there is the usual term $\Delta n_l(\x) = \delta_l(\x) (\partial n/\partial\delta_l)$ due to changes in mean
background density (as in the Gaussian case).

Putting this all together, we find that the long-wavelength halo density field in a $\gnl$ cosmology is 
given by:\footnote{Analogously to the $\fnl$ case, we have neglected two terms in Eq.~(\ref{eq:pbs_gnl}).
The term $\gnl(\Phi_l(\x)^3 - 3 \langle\Phi_l^2\rangle \Phi_l(\x))$
only alters power spectra at order $\bigoh(\gnl^2)$, and we will neglect terms of this order.
The term $\gnl(\Phi_s(\x)^3 - 3 \langle\Phi_s^2\rangle \Phi_s(\x))$ generates kurtosis in the density field and modifies
the halo mass function \cite{LoVerde:2011iz}, but in a way which is independent of $\Phi_l$ and therefore does not contribute
to large-scale clustering.}
\ba
n_l(\x) &=& {\bar n} 
    + \frac{\partial n}{\partial \delta_l} \delta_l(\x) 
    + 3 \gnl \frac{\partial n}{\partial\log\Delta_\Phi} (\Phi_l(\x)^2 - \langle \Phi_l^2 \rangle)
    + 3 \gnl \frac{\partial n}{\partial\fnl} \Phi_l(\x) \nn \\
  &=& {\bar n} \left( 1 + b_1 \delta_l(\x) + \frac{3}{2} \beta_f \gnl (\Phi_l(\x)^2-\langle\Phi_l^2\rangle) + \beta_g \gnl \Phi_l(\x) \right)  \label{eq:pbs_gnl_nl}
\ea
where $b_1$ and $\beta_f$ were defined previously (Eqs.~(\ref{eq:pbs_b0def}),~(\ref{eq:pbs_b1def})), and:
\be
\beta_g = 3 \frac{\partial\log n}{\partial\fnl}  \label{eq:pbs_b2def}
\ee
The large-scale halo bias $b(k) = P_{mh}(k)/P_{mm}(k)$ is given by:
\be
b(k) = b_1 + \frac{\beta_g \gnl}{\alpha(k,z)}   \label{eq:pbs_gnl_bias}\,.
\ee
Note that the $(\beta_f \gnl)$ term in Eq.~(\ref{eq:pbs_gnl_nl}) does not contribute to the bias,
since the field $(\Phi_l(\x)^2 - \langle \Phi_l^2 \rangle)$ and the long-wavelength
density field $\delta_l$ are uncorrelated (their cross correlation is a three-point
function of Gaussian fields, which vanishes).
This term should generate stochastic bias, but we defer a systematic study of halo stochasticity 
in non-Gaussian cosmologies to a future paper \cite{simone_stochasticity}.

We have now arrived at the peak-background split prediction~(\ref{eq:pbs_gnl_bias}) for halo bias in a $\gnl$ cosmology,
which relates the scale-dependent $\gnl$ bias to the derivative $(\partial\log n/\partial\fnl)$
of the halo mass function in an $\fnl$ cosmology.
In the terminology of the previous subsection, this is a ``weak'' prediction: we have shown that the problem of
computing the $\gnl$ bias is naturally related to the problem of understanding the mass function in an $\fnl$
cosmology, but the coefficient $\beta_g$ has not been expressed in terms of observable quantities.

To obtain a ``strong'' prediction, we need to evaluate the derivative $(\partial\log n/\partial\fnl)$, which
requires making additional assumptions.
This has been done in \cite{LoVerde:2011iz}, assuming a barrier crossing model
for the mass function and using the Edgeworth expansion to calculate the derivative 
(see also \cite{Matarrese:2000iz,LoVerde:2007ri,Chongchitnan:2010xz,Lam:2009nb,Maggiore:2009rx,DeSimone:2010mu,D'Amico:2010ta}).
The result is:
\be
\frac{\partial \log n(M)}{\partial\fnl} = \frac{\kappa_3(M)}{6} H_3(\nu(M)) - \frac{1}{6} \frac{d\kappa_3/dM}{d\nu/dM} H_2(\nu(M))  \label{eq:edgeworth_dndfnl}
\ee
where $\nu=\delta_c/\sigma_M$, and $H_2(x) = x^2-1$ and $H_3(x)=x^3-3x$ are Hermite polynomials.
We will compare this prediction with $N$-body simulations in \S\ref{sec:results}.

\section{Barrier crossing model}
\label{sec:barrier}

In this section, we will study large-scale bias using a barrier crossing model, obtaining
results which are consistent with the peak-background split formalism from the previous section.
The two approaches are complementary: the barrier model has the advantage that it generates complete predictions
for halo statistics (such as the mass function or bias) via an algorithmic calculational procedure, but obscures the
physical intuition of the peak-background split.
For completeness, the calculations in this section will be sufficiently general to include the
cases of Gaussian, $\fnl$-type, and $\gnl$-type initial conditions.

\subsection{Setting up the calculation}
\label{ssec:barrier_setup}

The barrier crossing model is an old, widely influential idea in cosmology, in which halos of mass $\ge M$ are identified
with peaks in the smoothed linear density field \cite{Press:1973iz}.
Although more complex versions have been proposed,
we will use the simplest version: a spherical collapse model with constant barrier height,
defined formally as follows.

We model halos of mass $\ge M$ as regions where the smoothed linear density field $\delta_M(\x)$ (defined in Eq.~(\ref{eq:deltaM_def}))
exceeds the threshold value $\delta_c$, i.e.~the halo number density $n_h(\x)$ is given by:
\be
n_h(\x) = \frac{\rho_m}{M} \theta(\delta_M(\x) - \delta_c)  \label{eq:barrier_nh}
\ee
where $\theta$ is the step function
\be
\theta(x) = \left\{
  \begin{array}{cl}
     0 & \mbox{if $x < 0$} \\
     1 & \mbox{if $x \ge 0$}
  \end{array}
\right.
\ee
Throughout this paper, we take $\delta_c=1.42$; this value produces somewhat improved agreement between the barrier
model and simulations, compared to the Press-Schechter value $\delta_c=1.69$.\footnote{We experimented with using 
a mass-dependent barrier $\delta_c(\nu)$ chosen for consistency with a universal
mass function such as Sheth-Tormen \cite{Sheth:1999mn} or Warren \cite{Warren:2005ey}, but found that this did not result
in further improvement.}

To study halo bias in this model, we define the following notation.
Let $\x$, $\x'$ be two points separated by distance $r$, let $\delta_{\rm lin}$ denote the (unsmoothed)
linear density field at $\x$, and let $\delta_M'$ denote the smoothed linear density field at $\x'$.
We denote the joint PDF of these random variables by $p(\delta_{\rm lin}, \delta_M')$, and denote the 1-variable
PDF of $\delta_M'$ by $p(\delta_M')$.
We define
\ba
p_0 &=& \int_{\delta_c}^\infty d\delta_M'\, p(\delta_M')  \label{eq:p0_def} \\
\xi_0(r) &=& \int d\delta_{\rm lin}\, d\delta_M'\,
   p(\delta_{\rm lin}, \delta_M')
   \, \delta_{\rm lin} \theta(\delta_M'-\delta_c)  \label{eq:xi0_def}
\ea
These quantities are related to the halo mass function $n(M)$ and matter-halo correlation function $\xi_{mh}(r)$, 
but there is one wrinkle.
In the barrier crossing model, the field $n_h$ defined in Eq.~(\ref{eq:barrier_nh}) represents the number density
of halos with mass $\ge M$, whereas we want to consider a sample of halos with mass in a narrow mass range $(M,M+dM)$.
Thus $n(M)$ and $\xi_{mh}(r)$ are obtained by taking derivatives as follows:
\ba
n(M) &=& -\frac{2\rho_m}{M} \left( \frac{dp_0}{dM} \right) \\
\xi_{mh}(r) &=& \frac{d\xi_0(r)/dM}{dp_0/dM}  \label{eq:xi_derivative}
\ea

\subsection{Mass function, halo bias, and interpretation}

In principle, calculation of the halo mass function and large-scale bias in the barrier crossing model
has now been reduced to evaluation of Eqs.~(\ref{eq:p0_def})--(\ref{eq:xi_derivative}).
We defer details of the calculation to Appendix~\ref{app:barrier} and quote the final results.
The halo mass function is given by:
\ba
n(M) &=& \frac{2\rho_m}{M} \left( \frac{d\log\sigma^{-1}}{dM} \right) \frac{e^{-\nu^2/2}}{(2\pi)^{1/2}}
   \Bigg[ \nu + \fnl \left( \kappa_3^{(1)}(M) \frac{\nu H_3(\nu)}{6} - \frac{d\kappa_3^{(1)}/dM}{d(\log\sigma^{-1})/dM} \frac{H_2(\nu)}{6} \right) \nn \\
   && \hspace{4cm} + \gnl \left( \kappa_4^{(1)}(M) \frac{\nu H_4(\nu)}{24} - \frac{d\kappa_4^{(1)}/dM}{d(\log\sigma^{-1})/dM} \frac{H_3(\nu)}{24} \right) \Bigg]  
     \label{eq:barrier_mf}
\ea
The halo bias $b(k) = P_{mh}(k)/P_{mm}(k)$ is given by (in the large-scale limit $k\rightarrow 0$):
\be
b(k) = b_1 + b_{1f} \fnl + b_{1g} \gnl + \frac{\beta_f \fnl + \beta_g \gnl}{\alpha(k)}    \label{eq:barrier_bias}
\ee
where:
\ba
b_1 &=& 1 + \frac{\nu^2-1}{\delta_c}  \label{eq:barrier_b0} \\
b_{1f} &=& -\kappa_3^{(1)}(M) \left( \frac{\nu^3-\nu}{2\delta_c} \right) 
  + \frac{d\kappa_3^{(1)}/dM}{d(\log\sigma^{-1})/dM} \left( \frac{\nu+\nu^{-1}}{6\delta_c} \right)  \label{eq:barrier_b0f} \\
b_{1g} &=& -\kappa_4^{(1)}(M) \left( \frac{\nu^4-3\nu^2}{6\delta_c} \right) 
  + \frac{d\kappa_4^{(1)}/dM}{d(\log\sigma^{-1})/dM} \left( \frac{\nu^2}{12\delta_c} \right)  \label{eq:barrier_b0g} \\
\beta_f &=& 2\nu^2-2   \label{eq:barrier_b1} \\
\beta_g &=& \kappa_3^{(1)}(M) \frac{\nu^3-3\nu}{2} - \frac{d\kappa_3^{(1)}/dM}{d(\log\sigma^{-1})/dM} \left( \frac{\nu-\nu^{-1}}{2} \right)  \label{eq:barrier_b2}
\ea
Although the above expressions are the result of a purely formal calculation, we will now show that each term has a natural
interpretation.

Considering first the halo mass function~(\ref{eq:barrier_mf}), we have found a Press-Schechter mass function (with $\delta_c=1.42$)
in the Gaussian case, plus first-order corrections in $\fnl$ and $\gnl$ which agree with those found in \cite{LoVerde:2007ri,LoVerde:2011iz}
using the Edgeworth expansion.
This agreement is expected since the two calculations are based on the same barrier crossing model.

Moving on to halo bias, 
in the Gaussian case, we predict that $b(k)$ is constant on large scales, with value $b_1$ given by Eq.~(\ref{eq:barrier_b0}).
The peak-background split argument suggests
a general relation between the large-scale halo bias and the halo mass function
which applies generally to a universal mass function of the form:
\be
n(M) = \frac{\rho_m}{M} f(\nu) \frac{d \log\sigma^{-1}}{dM}
\ee
On large scales, the bias is predicted to be scale-independent and given by \cite{Cole:1989vx}:
\be
b_1 = 1 - \frac{\nu}{\delta_c} \frac{d\log f}{d\nu}  \label{eq:pbs_bias_univ}
\ee
Comparing our predictions~(\ref{eq:barrier_mf}),~(\ref{eq:barrier_b0}) for $n(M)$ and $b_1$, we find agreement, 
i.e. Eq.~(\ref{eq:barrier_b0}) for $b_1$ can be interpreted as the general peak-background split expression
for halo bias, specialized to the Press-Schechter mass function.

More generally, the $b_{1f}$ and $b_{1g}$ contributions to the bias (Eqs.~(\ref{eq:barrier_b0f}),~(\ref{eq:barrier_b0g}))
represent shifts in the scale-independent part of the bias due to primordial non-Gaussianity.
It is straightforward to check that these terms can be obtained by plugging the non-Gaussian mass function in
Eq.~(\ref{eq:barrier_mf}) into the peak-background split prediction~(\ref{eq:pbs_bias_univ}) for scale-independent bias,
i.e.~the $b_{1f}$ and $b_{1g}$ terms can be interpreted as changes to the bias which are entirely due to the mass function
being perturbed in a non-Gaussian cosmology.
This type of term (scale-independent bias proportional to $\fnl$) was first found for $\fnl$ cosmologies in \cite{Desjacques:2008vf}.
Note that a scale-independent shift is unobservable in practice, and cannot be used to constrain non-Gaussianity,
since the bias of a real tracer population, such as galaxies or quasars, is a free parameter.

The $\beta_f$ contribution to the bias is the well-known scale-dependent bias in an $\fnl$ cosmology.  Comparing Eq.~(\ref{eq:barrier_b1})
for $\beta_f$ with Eqs.~(\ref{eq:barrier_mf}),~(\ref{eq:barrier_b0}), this term can be written either as $\beta_f = 2\partial(\log n)/\partial(\log\Delta_\Phi)$
or $\beta_f = 2\delta_c(b_1-1)$.
(In \S\ref{sec:pbs}, we referred to these as ``weak'' and ``strong'' predictions.)

The $\beta_g$ contribution to the bias is the focus of this paper: scale-dependent bias in a $\gnl$ cosmology.
Eq.~(\ref{eq:barrier_b2}) gives this term in the ``strong'' form that was found previously (Eq.~(\ref{eq:edgeworth_dndfnl})) 
using the peak-background split argument.
Alternately, we can write this term in the ``weak'' form $\beta_g = 3\partial(\log n)/\partial\fnl$ using Eq.~(\ref{eq:barrier_mf}).

In summary, we have found that the complete expression for large-scale halo bias in the barrier crossing model (Eq.~(\ref{eq:barrier_bias}))
agrees perfectly with the peak-background split calculation from \S\ref{sec:pbs}.
The bias contains a scale-independent part $(b_1 + b_{1f}\fnl + b_{1g}\gnl)$ which can be obtained from the halo mass function,
via the general relation~(\ref{eq:pbs_bias_univ}).
The scale-independent bias depends on $\fnl$ and $\gnl$, because the halo mass function depends on these parameters.
The bias also contains a scale-dependent part $(\beta_f \fnl + \beta_g\gnl) / \alpha(k)$ whose coefficients can be calculated explicitly
and agree with the peak-background split predictions.

\subsection{Comparison with previous work}

It is interesting to compare the above calculations with the results of \cite{Desjacques:2009jb} 
(see also \cite{Giannantonio:2009ak}), where $\beta_g$ was calculated using the MLB formula \cite{Matarrese:1986et},
which gives $N$-point functions of halos as an asymptotic series in $\nu$.
The scale-dependent $\gnl$ bias was found to be (in our notation):
\be
\beta_g^{\rm MLB} = \kappa_3^{(1)}(M) \frac{\delta_c \nu (b_1-1)}{2}
\ee
When this prediction was compared to $N$-body simulations, it was found to be a poor fit.

Comparing $\beta_g^{\rm MLB}$ with our calculation~(\ref{eq:barrier_b2}) for $\beta_g$, it is seen that
the two agree in the high-mass limit $\nu\rightarrow\infty$, but disagree in subleading terms.
This is expected since the MLB formula is based on the same barrier crossing model that we have used,
but it is an asymptotic result, whereas we have done an exact calculation (to first order in $\fnl$,
$\gnl$).
For realistic halo masses, the ``subleading'' terms neglected in the MLB formula are of order one (to quantify this better,
$\beta_g$ and $\beta_g^{\rm MLB}$ agree to 10\% only when the halo bias $b_1 \ge 15$), so in practice the two predictions
are quite different.

Recently, ref.~\cite{Desjacques:2011mq} argued that the barrier crossing model cannot generate
correct predictions for general non-Gaussian initial conditions such as the $\gnl$ model, but we found
the opposite conclusion: brute-force calculation in the barrier crossing model, collecting all
terms of order $\bigoh(\gnl)$, agrees precisely (i.e.~to all orders in $1/\nu$) with the peak-background split.
It seems intuitively plausible that two must be consistent, since the peak-background split argument depends
only on the assumption that halo formation is determined by the local density field, and the barrier crossing 
model is a concrete example of a model in which this assumption is satisfied.

\section{Results from $N$-body simulations}
\label{sec:results}

In the last two sections, we have obtained complete analytic predictions for large-scale bias in
a $\gnl$ cosmology, finding agreement between the peak-background split formalism (\S\ref{sec:pbs})
and a barrier crossing model based on spherical collapse (\S\ref{sec:barrier}).

To compare these predictions with simulation, we
performed collisionless $N$-body simulations using the GADGET-2 TreePM code \cite{Springel:2005mi}.
Simulations were done using periodic
box size $R_{\rm box} = 1600$ $h^{-1}$~Mpc, particle count $N_p = 1024^3$, and force softening
length $R_s = 0.05 (R_{\rm box}/N_p^{1/3})$.
With these parameters and the fiducial cosmology from \S\ref{sec:intro}, the particle mass 
is $m_p = 2.92 \times 10^{11}$ $h^{-1}$~$M_\odot$.

We generate initial conditions by simulating a Gaussian primordial potential $\Phi$, and applying
$\fnl$ or $\gnl$ corrections by straightforward use of Eq.~(\ref{eq:local_ng}).
We linearly evolve to redshift $z_{\rm ini}=100$ using the transfer function\footnote{One subtlety here:
straightforward use of CAMB's transfer function at redshift 100 leads to inconsistencies since CAMB includes
radiation (which is not negligible at $z=100$) in its expansion history, while GADGET does not.  For this
reason we use CAMB's linear transfer function at low redshift and extrapolate back to $z=100$ using the growth
function in an $\Omega_{\rm rad}=0$ universe.} from CAMB \cite{Lewis:1999bs},
and obtain initial particle positions at this redshift using the Zeldovich approximation \cite{Zeldovich:1969sb}.
(At $z_{\rm ini}=100$, transient effects due to use of this approximation should be negligible \cite{Crocce:2006ve}.)

After running the $N$-body simulation, we group particles into halos using an MPI parallelized implementation of the 
friends-of-friends algorithm \cite{Frenk:1988zz}
with link length $L_{\rm FOF} = 0.2 R_{\rm box} N_p^{-1/3}$.
For a halo containing $N_{\rm FOF}$ particles, we assign a halo position given by the mean of the individual
particle positions.
We estimate halo bias $b(k) = P_{mh}(k)/P_{mm}(k)$ using the procedure described in Appendix~A of \cite{Smith:2010gx}.
The statistical error $\Delta b(k)$ obtained using this procedure is smaller than the error that would be obtained
assuming uncorrelated estimates of the power spectra $P_{mm}$ and $P_{mh}$, since shared sample variance is taken
into account.

Results in this paper are based on 4 simulations with Gaussian initial conditions,
5 simulations with $\gnl=\pm 2\times 10^6$, and 3 simulations with $\fnl=\pm 250$
(for a total of 20 simulations).

\subsection{Fitting the functional form $b(k) = b_1 + \beta_g\gnl/\alpha(k)$}
\label{ssec:bias_fits}

We now compare our analytic prediction for $b(k)$ to simulation in several steps, corresponding to increasingly strong 
versions of the prediction.

\begin{figure}
\centerline{\includegraphics[width=12cm]{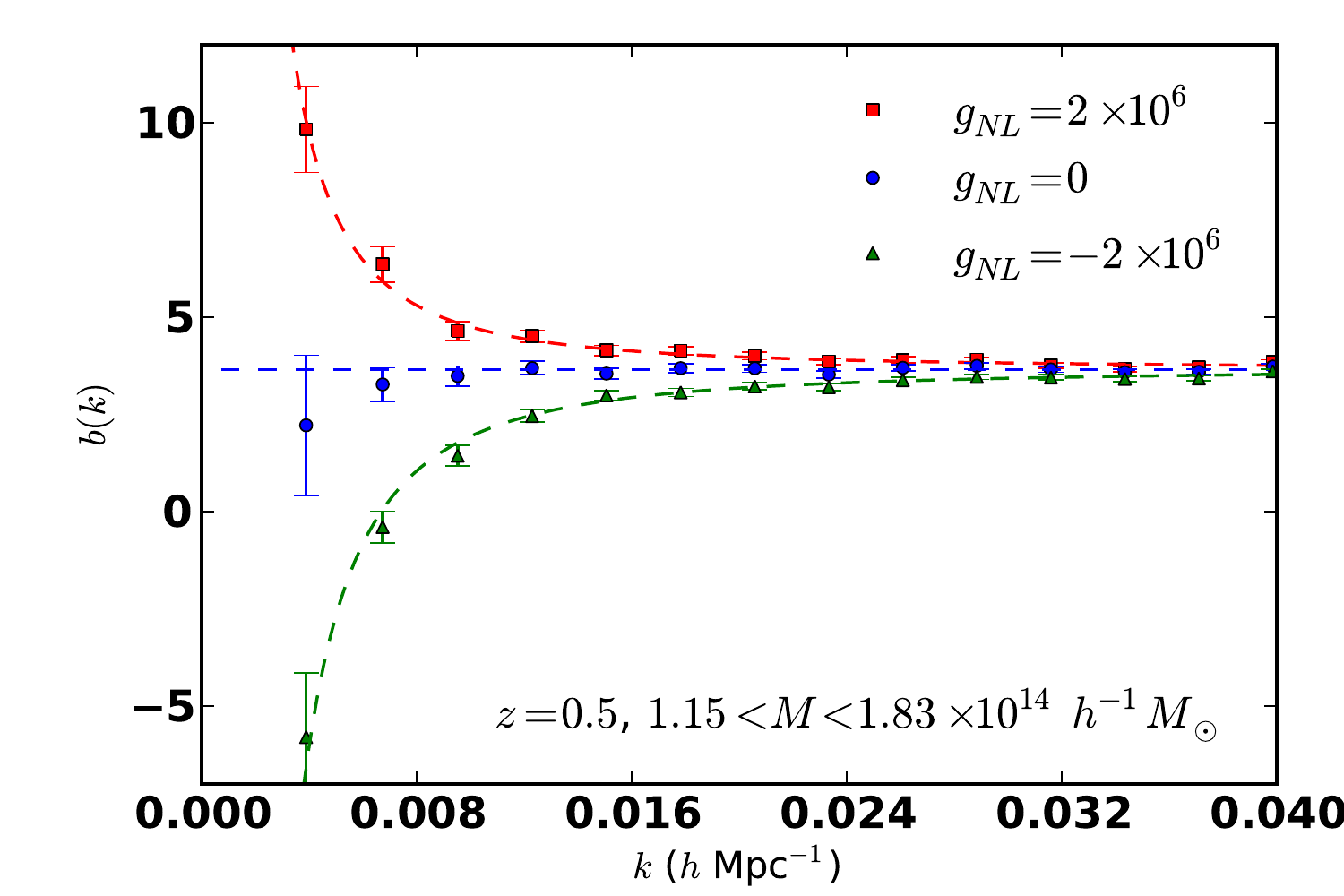}}
\caption{An example to illustrate that halo bias in a $\gnl$ cosmology takes the functional form
form $b(k) = b_1 + \beta_g \gnl/\alpha(k)$.  This figure corresponds to redshift $z=0.5$ and halo
mass range $1.15 \le M \le 1.83 \times 10^{14}$ $h^{-1}$ $M_\odot$, but we find the same functional
form for all redshifts and halo masses.}
\label{fig:example_fit}
\end{figure}

First, consider the weakest possible question: our analytic prediction for the bias is of the functional form
\be
b(k) = b_1 + \beta_g \frac{\gnl}{\alpha(k)}  \label{eq:b0b2_fit}
\ee
Is this is a good fit to simulation, if we treat the coefficients $b_1$ and $\beta_g$ as free parameters?
(We will compare our analytic prediction for $\beta_g$ to simulation in the next subsection;
for now we are just testing whether the functional form~(\ref{eq:b0b2_fit}) is correct.)

In Fig.~\ref{fig:example_fit}, we show some example fits of this form, for redshift
$z=0.5$ and halo mass range $1.15 \le M \le 1.83\times 10^{14}$ $h^{-1}$ $M_\odot$.
Each fit was performed using bias estimates from 4 independent simulations with $L_{\rm box} = 1600$ $h^{-1}$ Mpc
and wavenumbers $k \le 0.04$ $h$ Mpc$^{-1}$.
We find good $\chi^2$ values for the fits, with recovered parameters:
\be
\begin{array}{cc}
b_1 = 3.653 \pm 0.026 & \mbox{\hspace{1cm} for $\gnl=0$} \\
(b_1,10^3\beta_g)=(3.575 \pm 0.038, 0.581\pm 0.056) & \mbox{\hspace{1cm} for $\gnl=2\times 10^6$} \\
(b_1,10^3\beta_g)=(3.824 \pm 0.039, 0.935 \pm 0.060) & \mbox{\hspace{1cm} for $\gnl=-2\times 10^6$}
\end{array}
\label{eq:example_b2_fit}
\ee
We note that the recovered bias parameters~(\ref{eq:example_b2_fit}) in this example show that both $b_1$ and $\beta_g$ are $\gnl$-dependent.
In the barrier crossing model, we made a prediction for the $\gnl$ dependence of $b_1$ (Eq.~(\ref{eq:barrier_b0g})).
We find good agreement between this prediction and our simulations.
Note that in practice, the $\gnl$ dependence of $b_1$ is unobservable since for a real tracer population, the halo occupation
distribution is not known precisely and $b_1$ must be treated as a free parameter to be determined from data.

The observed $\gnl$ dependence of $\beta_g$ corresponds to scale-dependent bias of order $\bigoh(\gnl^2)$ or higher (note that $\beta_g$
is defined in such a way that constant $\beta_g$ corresponds to scale-dependent bias which is linear in $\gnl$).
This complicates comparison with our analytic predictions, since we have only calculated the bias to order
$\bigoh(\gnl)$.
We address this by estimating $\beta_g$ by averaging the estimates obtained
from simulations with $\gnl=\pm 2\times 10^6$, thus removing contributions to $b(k)$ which are proportional to $\gnl^2$.
Note that this does not remove $\bigoh(\gnl^3)$ contributions to the bias,
but we have checked that such contributions are negligible for $\gnl=\pm 2\times 10^6$, by comparing with simulations
with halved step size.

Repeating this fitting procedure for redshifts $z\in\{2,1,0.5,0\}$ and a range of halo masses (the precise set of
halo mass bins used is shown in Fig.~\ref{fig:b2_comparison} below), we find $\chi^2$ values which are consistent with their
expected distribution, i.e.~we find that the functional form~(\ref{eq:b0b2_fit}) is a good fit to the simulations 
for a wide range of redshifts and halo masses.
For this reason, in subsequent sections, we will ``compress'' the estimates of $b(k)$ in each simulation (as shown
in Fig.~\ref{fig:example_fit}) to two numbers ($b_1$ and $\beta_g$), with statistical errors given by the fitting
procedure.

\subsection{Comparison with analytic predictions}

Now that we have established the functional form $b(k)=b_1 + \beta_g\gnl/\alpha(k)$ of the bias,
and a procedure for estimating $\beta_g$ from simulation as a function of redshift
and halo mass, we would like to compare with our analytic predictions for $\beta_g$.

\begin{figure}
\centerline{\includegraphics[width=9cm]{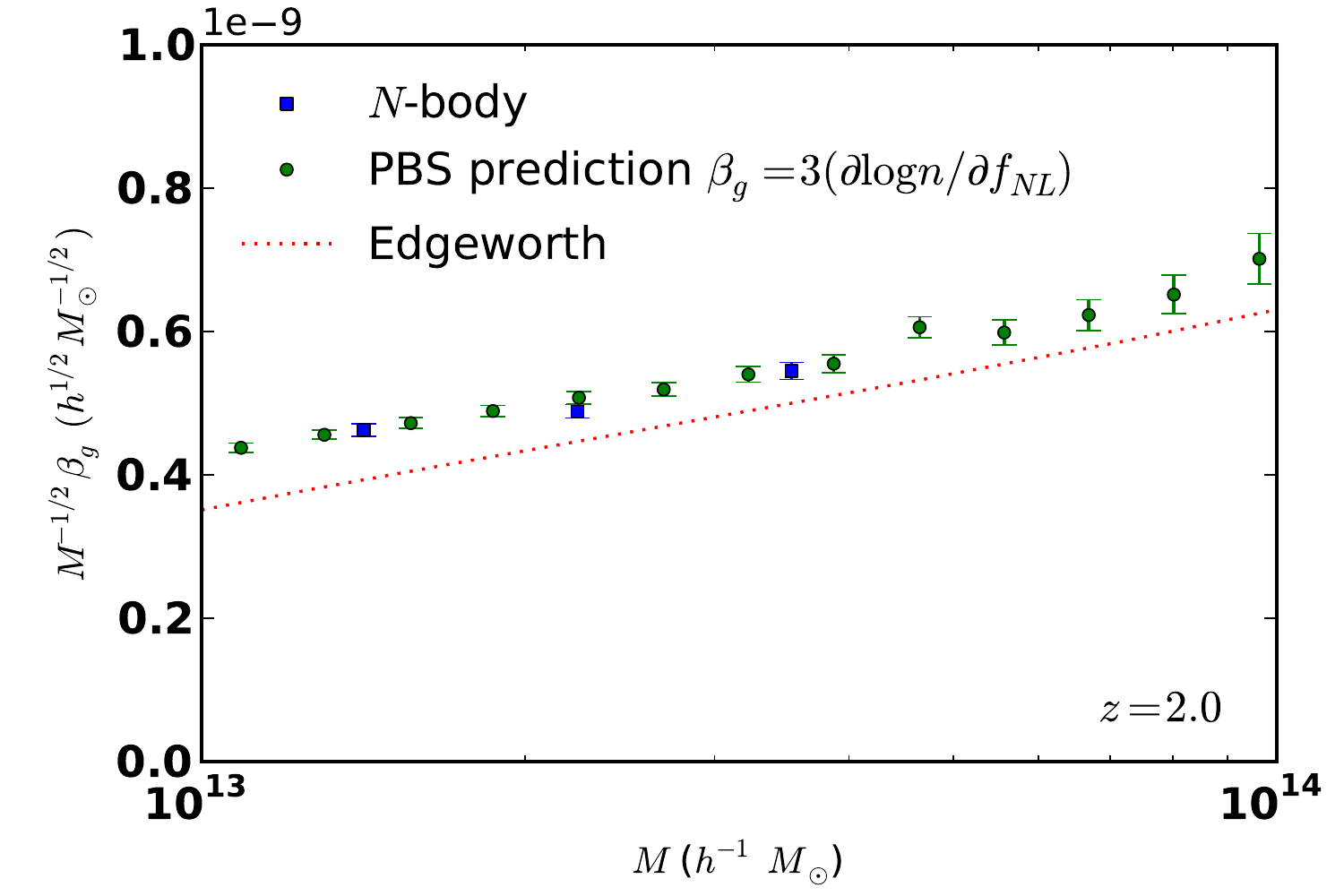} \includegraphics[width=9cm]{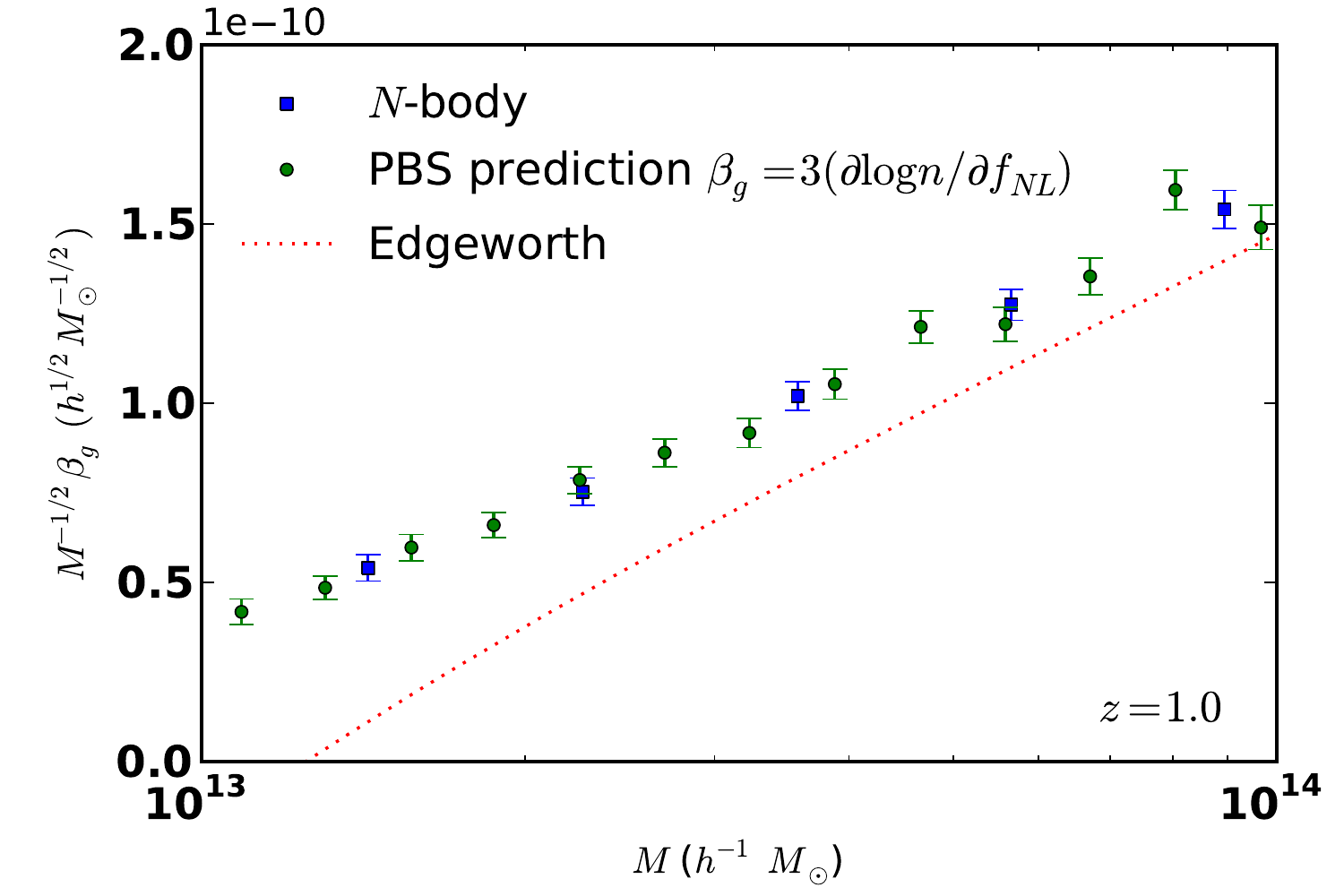}}
\centerline{\includegraphics[width=9cm]{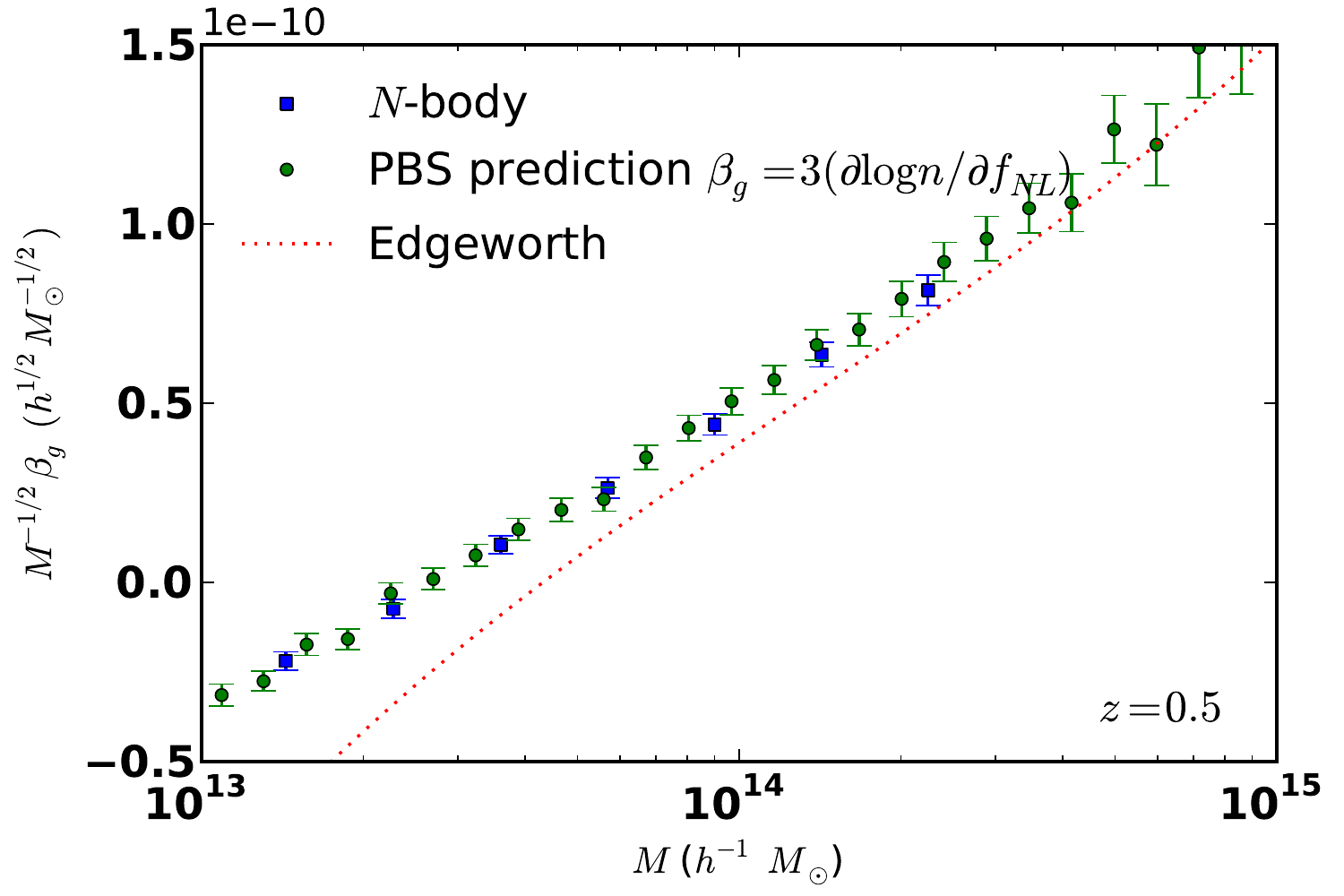} \includegraphics[width=9cm]{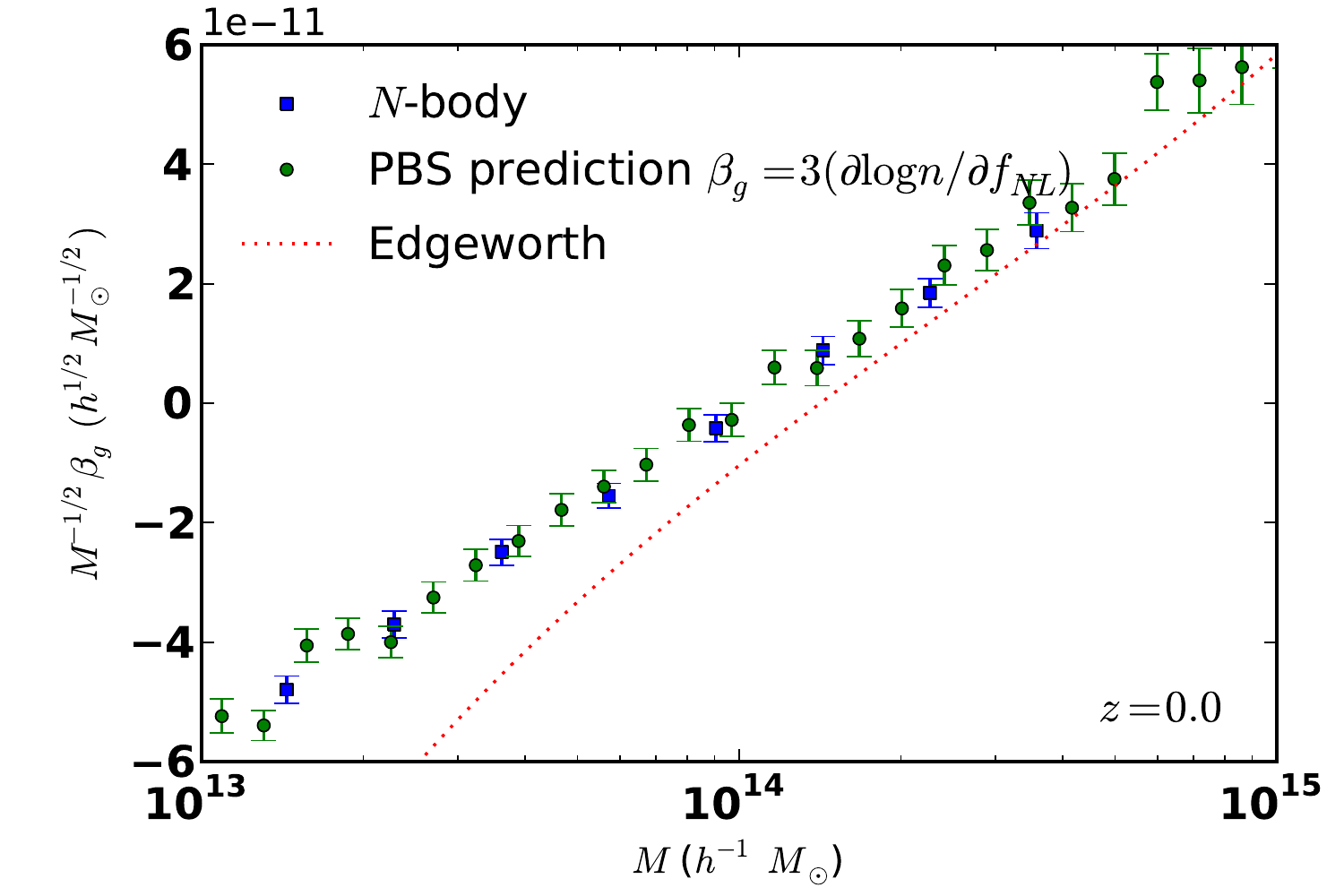}}
\caption{Comparison of  the ``weak" and ``strong" predictions for the scale-dependent bias in a $\gnl$ cosmology.
{\bf Blue squares:} Direct estimates of the bias, extracted from simulations with $\gnl = \pm 2 \times 10^6$
as described in \S\ref{ssec:bias_fits}.
{\bf Green circles:} ``Weak'' analytic prediction for the bias ($\beta_g = 3(\partial\log n/\partial\fnl)$)
from the peak-background split formalism, showing perfect agreement.
The estimates of $(\partial\log n/\partial\fnl)$ shown in the figure were obtained directly from simulations
with $\fnl=\pm 250$.
{\bf Red dotted curve:} Edgeworth prediction for the bias (Eq.~(\ref{eq:barrier_b2})).  Good agreement
is seen at high mass, but at low masses Edgeworth underpredicts $3(d\log n/d\fnl)$.
We will find an improvement in \S\ref{ssec:bias_final}.}
\label{fig:b2_comparison}
\end{figure}

First, consider the ``weak'' form of the prediction ($\beta_g = 3 (\partial\log n/\partial\fnl)$)
obtained from the peak-background split argument.
We can test this prediction by estimating the derivative $(\partial\log n/\partial\fnl)$ directly
from simulations, by taking finite differences of $\log(n)$ in simulations
with $\fnl=\pm 250$.  (We checked convergence in the step size.)
We find that the prediction holds perfectly (within the statistical errors of the simulations)
for all redshifts and halo masses (Fig.~\ref{fig:b2_comparison}).

Second, consider the ``strong'' Edgeworth prediction (Eq.~(\ref{eq:barrier_b2})), in which an
explicit formula for $\beta_g$ is given.
In this case, we find reasonable agreement at high mass ($M \gsim 10^{14}$ $h^{-1}$ $M_\odot$), 
but the prediction breaks down at low halo mass (Fig.~\ref{fig:b2_comparison}).

Our interpretation is as follows.
The peak-background split prediction $\beta_g = 3(\partial\log n/\partial\fnl)$ is a universal
relation between bias in a $\gnl$ cosmology and the mass function in an $\fnl$ cosmology.
Although ``weak'' in the sense that it does not supply a closed-form expression for $\beta_g$,
the derivation makes few assumptions, and one expects it to be exact.
In order to constrain $\gnl$ from real data, we need a ``strong'' prediction which expresses
$\beta_g$ in closed form, using only observable quantities (i.e.~the analog of the Dalal et
al formula $\beta_f = 2 \delta_c (b_1-1)$ for an $\fnl$ cosmology).
Using the Edgeworth expansion, 
one can make such a prediction in the context of the barrier crossing model (Eq.~(\ref{eq:barrier_b2})), and obtain rough agreement
with simulations, but the level of agreement is not really good enough for doing precision
cosmology.
Therefore, we next propose a slightly modified version of the Edgeworth prediction.

\subsection{A simple universal formula for the bias in a $\gnl$ cosmology}
\label{ssec:bias_final}

We would like to slightly modify the Edgeworth prediction~(\ref{eq:barrier_b2}) for $\beta_g$
so that it agrees better with $N$-body simulations.
It is also convenient to have a prediction in which $\beta_g$ is given as a function of
observable quantities:  Gaussian bias $b_1$ (rather than halo mass, which is unobservable)
and redshift $z$.

\begin{figure}
\centerline{\includegraphics[width=12cm]{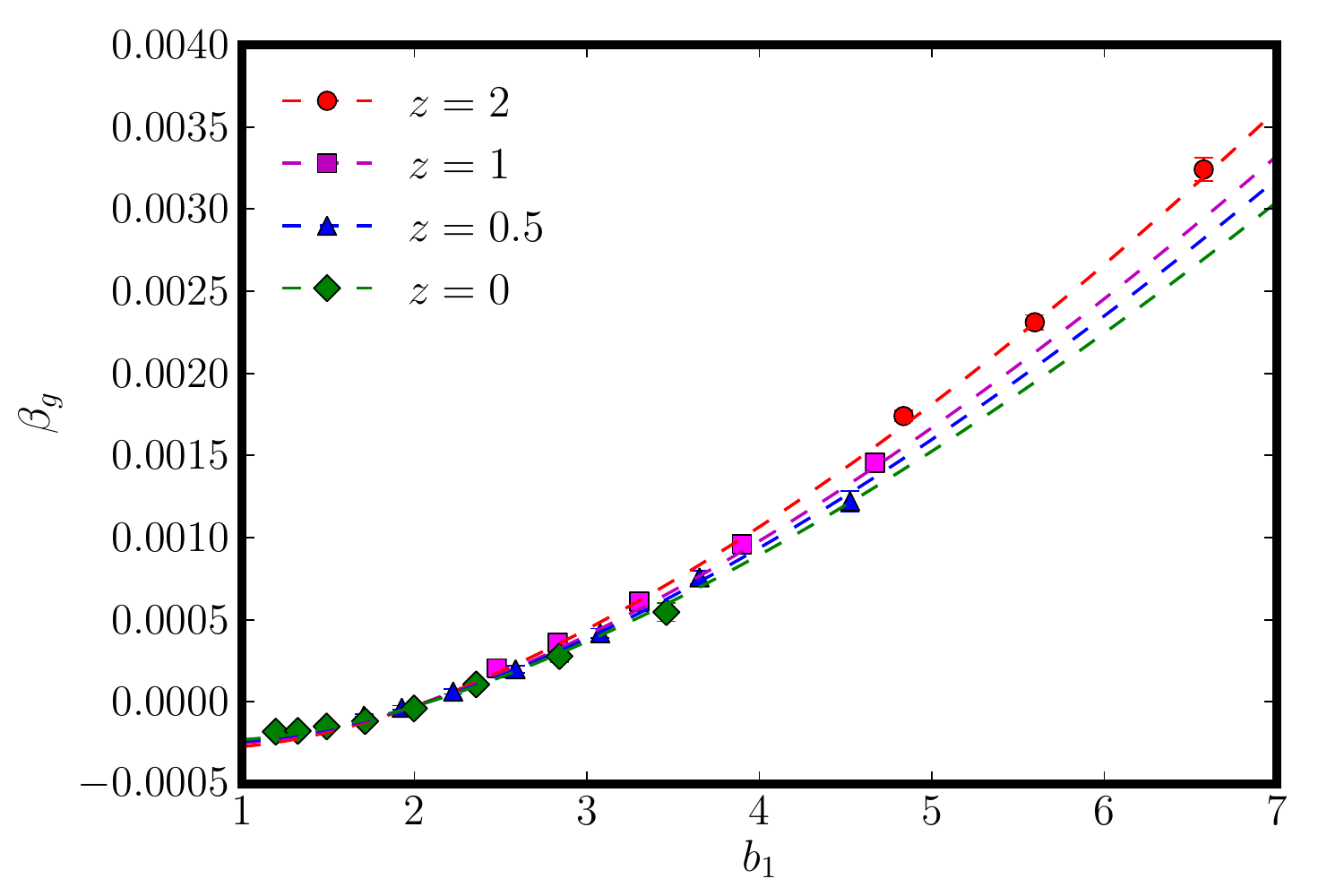}}
\caption{Scale-dependent $\gnl$ bias coefficient $\beta_g$ as a function of redshift $z$ and
halo bias $b_1$, showing excellent agreement between our final analytic result (Eq.~(\ref{eq:b2_final}),
dashed curves) and $N$-body simulations (error bars).}
\label{fig:b2_vs_b0}
\end{figure}

We start by rewriting the Edgeworth prediction~(\ref{eq:barrier_b2}) for $\beta_g$ in terms of variables $(b_1,z)$.
The following fitting functions for $\kappa_3$ and $d\kappa_3/d\log(\sigma^{-1})$ are convenient:
\ba
\kappa_3 &=& 0.000329 (1 + 0.09z) b_1^{-0.09}  \label{eq:k3_bz} \\
\frac{d\kappa_3}{d\log\sigma^{-1}} &=& -0.000061 (1 + 0.22z) b_1^{-0.25} \label{eq:dk3_bz}
\ea
For purposes of this subsection, we {\em define} the quantity $\nu$ to be given in terms of $b_1$ and $z$ by:
\be
\nu = [\delta_c(b_1-1) + 1]^{1/2}  \hspace{1cm} \mbox{(where $\delta_c=1.42$)}  \label{eq:nu_bz}
\ee
The Edgeworth prediction for $\beta_g$ can be written in the following form:
\be
\beta_g^{\rm Edge.} = \kappa_3 \bigg[-1 + \frac{3}{2}(\nu-1)^2 + \frac{1}{2}(\nu-1)^3 \bigg]
  - \frac{d\kappa_3}{d\log\sigma^{-1}} \left( \frac{\nu-\nu^{-1}}{2} \right)  \label{eq:b2_pretweak}
\ee
Empirically, we find that if we tweak the Edgeworth prediction by changing the coefficients of the
polynomial in brackets as follows:
\be
\beta_g = \kappa_3 \bigg[ -0.7 + 1.4(\nu-1)^2 + 0.6(\nu-1)^3 \bigg]
  - \frac{d\kappa_3}{d\log\sigma^{-1}} \left( \frac{\nu-\nu^{-1}}{2} \right)  \label{eq:b2_final}
\ee
then we obtain good agreement with simulations (Fig.~\ref{fig:b2_vs_b0}).
The expression~(\ref{eq:b2_final}) for $\beta_g$ (with quantities $\kappa_3$, $d\kappa_3/d\log\sigma^{-1}$, $\nu$
defined by Eqs.~(\ref{eq:k3_bz})--(\ref{eq:nu_bz}))
is one of the main results of this paper and is our observational ``bottom line''
when constraining $\gnl$ from real data.

We have motivated our ``tweak'' to the Edgeworth prediction as essentially a fitting function
for the $\nu$ dependence (although it is worth noting that the $z$ dependence is correctly predicted
by the barrier crossing model).
A speculative interpretation of this tweak, which we will defer for future work, is as follows.
In the barrier crossing model, the second-order halo bias is given by $b_2 = (\nu^3-3\nu)/(\delta_c \sigma_M)$.
It is tempting to conjecture that the expression in brackets in Eq.~(\ref{eq:b2_final}) is
generally equal to $(\delta_c \sigma_M b_2)$, and interpret our ``tweak'' to the Edgeworth prediction~(\ref{eq:b2_pretweak})
as perturbing the relation between $b_1$ and $b_2$, relative to the barrier crossing model.
This opens up the possibility of directly measuring the second-order bias and determining $\beta_g$ directly.
To study the viability of this idea, one would need to compare $\beta_g$ in simulation to some other estimate
of second-order halo bias, such as the halo bispectrum in the squeezed limit.

\subsection{An important caveat}
\label{ssec:caveat}

There is an important caveat when using Eq.~(\ref{eq:b2_final}), or indeed any fitting function for
the $\gnl$ bias, to constrain $\gnl$ from real data.
It is tempting to compute $\beta_g$ by simply plugging the observed 
bias $b_1$ and redshift $z$ into Eq.~(\ref{eq:b2_final}).  (Since the $z$-dependence is very mild, a rough estimate
for the redshift suffices.)  However, we have only shown that this procedure is correct in the limit of a narrow bin
in halo mass and redshift, and a real tracer population will be a weighted average over $M$ and $z$.  

For example, consider the case in which the ``tracers'' are the dark matter particles themselves, i.e.~each
halo is weighted in proportion to its mass (assuming all mass is in halos).
This tracer population has bias $b_1=1$ (for the trivial reason that we are back to the dark matter field), so straightforward
use of Eq.~(\ref{eq:b2_final}) would suggest that $\beta_g \approx -0.00025$.  (This value would make the low-$k$ power spectrum a 
reasonably sensitive probe of $\gnl$.)
In fact, the true $\beta_g$ of this tracer is zero, since the matter power spectrum $P_{mm}(k)$ does not contain a term
proportional to $\gnl/\alpha(k)$.
This example shows that the true $\gnl$ bias of a tracer population can
differ significantly from the value obtained by straightforward use of Eq.~(\ref{eq:b2_final}).
In general, the $\gnl$ bias will depend on the full HOD (halo occupation distribution) of the tracer
population, not only on the Gaussian bias $b_1$.\footnote{Note that there is no analogous caveat in
the $\fnl$ case.  Because the relation $\beta_f = 2\delta_c(b_1-1)$ is linear, it applies to both
a tracer population which is narrowly selected in $(M,z)$ and to a population which is an arbitrary
weighted average over $(M,z)$.}

One popular approach to modeling the HOD is to assume that halos below some minimum mass $M_{\rm min}$
do not host tracers, whereas the mean number of tracers in a halo of mass $M \ge M_{\rm min}$ is proportional
to the total mass $M$.
For reference, we give a fitting function for the $\gnl$ bias for this HOD:
\be
\beta_g = \kappa_3 \bigg[ -0.4(\nu-1) + 1.5(\nu-1)^2 + 0.6(\nu-1)^3 \bigg]  \label{eq:b2_mw}
\ee
where for purposes of this equation, $\kappa_3$ and $\nu$ are defined as functions of the observables
$b_1$ and $z$ by Eqs.~(\ref{eq:k3_bz}),~(\ref{eq:nu_bz}) above.

Eq.~(\ref{eq:b2_mw}) applies to a mass-weighted population of halos above $M_{\rm min}$, whereas Eq.~(\ref{eq:b2_final})
applies to a population which is narrowly selected in mass.
The two agree for $b_1 \gsim 2.5$, suggesting that HOD dependence is small in practice for highly biased samples,
but disagree qualitatively for $b_1 \lsim 2.5$.
For example, the $\gnl$ bias $\beta_g$ changes sign at $b_1 \approx 2.1$ for the narrowly selected sample (Eq.~(\ref{eq:b2_final})),
whereas $\beta_g$ is always positive for the mass-weighted sample (Eq.~(\ref{eq:b2_mw})).

Our perspective is that, in order to obtain $\gnl$ constraints which are robust to HOD modeling uncertainty,
one should use highly biased samples $(b_1\gsim 2.5)$, where this uncertainty will be minimized.
Samples which are not highly biased do not give robust constraints;
for example, a tracer population with $b_1=1.8$ can have a $\gnl$ bias $\beta_g$ which is negative, zero, or
positive, depending on the HOD.

For highly biased samples, it is useful to make the following observation:
the $\gnl$ bias $\beta_g^{\rm fit}$ which is obtained from straightforward use of Eq.~(\ref{eq:b2_final})
is always less than the true $\gnl$ bias $\beta_g^{\rm true}$.\footnote{This 
statement assumes
that the probability that a halo hosts a tracer is a function only of the mass and redshift.
If the probability depends strongly on additional variables such as merger history, triaxiality, etc.
then this will generate additional contributions to $\beta_g$, in analogy to the $\fnl$ case 
\cite{Slosar:2008hx,Reid:2010vc}.
In principle, selection biases to $\beta_g$ can be addressed by folding the selection into the mass function
when computing $\partial(\log n)/\partial \fnl$, but detailed study is beyond the scope of this paper.}
This follows from positivity of the second derivative $d^2\beta_g/db_1^2$.
It follows that a $\gnl$ constraint obtained using $\beta_g^{\rm fit}$ is always valid, but slightly
overestimates the statistical error that could be obtained if $\beta_g^{\rm true}$ were known.
This
effectively treats HOD uncertainty as an extra source of systematic error.

\section{Discussion}
\label{sec:discussion}

We have computed large-scale halo bias for non-Gaussian initial conditions,
using two analytic frameworks: the
peak-background split formalism (\S\ref{sec:pbs}) and a barrier crossing model (\S\ref{sec:barrier}),
finding agreement between the two.
Although our emphasis has been on the constant-$\fnl$ and constant-$\gnl$ models,
our calculational machinery should apply to more general non-Gaussian initial conditions.

The peak-background split formalism is simpler and also suggests a simple physical
picture of non-Gaussian cosmologies on large scales. 
In an $\fnl$ cosmology, the amplitude $\Delta_\Phi$
of the initial fluctuations is not spatially constant, but is proportional to $(1 + 2 \fnl\Phi_l)$.
Thus, $\Delta_\Phi$ has fluctuations on large scales
which are 100\% correlated with the long-wavelength potential, generating halo bias of the
form $(\beta_f \fnl/\alpha(k))$.  In a $\gnl$ cosmology, the small-scale skewness is nonzero and proportional 
to $(\gnl\Phi_l)$, leading to halo bias of the form $(\beta_g \gnl/\alpha(k))$.
The peak-background split argument is very useful for generating universal relations such as
$\beta_g = 3 \partial(\log n)/\partial\fnl$, which are ``weak'' in the sense that the RHS has
not been expressed in terms of observable quantities,
but have the advantage of being exact (as can be seen by comparing the
two sets of errorbars in Fig.~\ref{fig:b2_comparison}).

The barrier crossing model generates all terms in the 
large-scale bias, including terms such as $b_{1f}$ and $b_{1g}$ which are easy to miss, by a 
purely algorithmic calculational procedure.
In addition, the barrier crossing model generates ``strong'' forms of the bias coefficients
(e.g.~the Edgeworth expression~(\ref{eq:barrier_b2}) for $\beta_g$), which are closed-form
expressions in $M$ and $z$.
However, these expressions are not exact
because the barrier crossing model is approximation to the true process of halo formation.

To obtain a ``bottom line'' expression for the scale-dependent $\gnl$ bias $\beta_g$ in terms of
redshift $z$ and Gaussian bias $z$, we found it necessary to tweak slightly the $b_1$ dependence 
of the Edgeworth prediction, arriving at the expression~(\ref{eq:b2_final}) which agrees very
well with simulations.
The caveat is that Eq.~(\ref{eq:b2_final}) applies only to a halo population which has been
selected in a narrow halo mass and redshift range.
In principle, one can calculate $\beta_g$ for a tracer population by multiplying by the halo occupation
distribution and integrating over mass and redshift.
In practice, the HOD is not known precisely and we have argued in \S\ref{ssec:caveat} that
the best approach is to only use highly biased populations ($b \gsim 2.5$) for constraining $\gnl$.
Since $\beta_g$ is a rapidly increasing function of $b_1$, this strategy makes sense both from the
perspective of minimizing statistical errors, and systematic errors due to HOD uncertainty.
In data analysis, it may be useful to impose cuts which increase the mean halo bias at the
expense of reducing the number of tracers.
Another advantage of subdividing tracer populations is that this may permit $\fnl$ and $\gnl$
to be constrained simultaneously (with a single tracer population, the two are degenerate).

\subsection*{Acknowledgements}

We thank Neal Dalal, Vincent Desjacques, Chris Hirata, Donghui Jeong, Brant Robertson, Fabian Schmidt,
Neelima Sehgal, David Spergel and Matias Zaldarriaga for helpful discussions.
K.~M.~S.~is supported by a Lyman Spitzer fellowship in the Department of Astrophysical Sciences at Princeton University.
M.~L.~is supported as a Friends of the Institute for Advanced Study Member and by the NSF though AST-0807444.
S.~F.~is supported by the Martin Schwarzschild Fund in Astronomy at Princeton University.
Simulations in this paper were performed at the TIGRESS high performance computer center at Princeton University which is jointly 
supported by the Princeton Institute for Computational Science and Engineering and the Princeton University Office of Information Technology.

\bibliographystyle{utphysx}
\bibliography{gnl}

\providecommand{\href}[2]{#2}\begingroup\raggedright\begin{thebibliography}{10}

\bibitem{Komatsu:2010fb}
E.~Komatsu {\em et al.}, ``{Seven-Year Wilkinson Microwave Anisotropy Probe
  (WMAP) Observations: Cosmological Interpretation},''
\href{http://arxiv.org/abs/1001.4538}{{arXiv:1001.4538 [astro-ph.CO]}}.

\bibitem{Percival:2009xn}
{ SDSS} Collaboration, W.~J. Percival {\em et al.}, ``{Baryon Acoustic
  Oscillations in the Sloan Digital Sky Survey Data Release 7 Galaxy Sample},''
  \href{http://dx.doi.org/10.1111/j.1365-2966.2009.15812.x}{{\em Mon. Not. Roy.
  Astron. Soc.} {\bfseries 401} (2010) 2148--2168},
\href{http://arxiv.org/abs/0907.1660}{{arXiv:0907.1660 [astro-ph.CO]}}.

\bibitem{Reid:2009xm}
B.~A. Reid {\em et al.}, ``{Cosmological Constraints from the Clustering of the
  Sloan Digital Sky Survey DR7 Luminous Red Galaxies},''
  \href{http://dx.doi.org/10.1111/j.1365-2966.2010.16276.x}{{\em Mon. Not. Roy.
  Astron. Soc.} {\bfseries 404} (2010) 60--85},
\href{http://arxiv.org/abs/0907.1659}{{arXiv:0907.1659 [astro-ph.CO]}}.

\bibitem{Riess:2011yx}
A.~G. Riess {\em et al.}, ``{A 3\% Solution: Determination of the Hubble
  Constant with the Hubble Space Telescope and Wide Field Camera 3},''
  \href{http://dx.doi.org/10.1088/0004-637X/730/2/119}{{\em Astrophys. J.}
  {\bfseries 730} (2011) 119},
\href{http://arxiv.org/abs/1103.2976}{{arXiv:1103.2976 [astro-ph.CO]}}.

\bibitem{Kessler:2009ys}
R.~Kessler {\em et al.}, ``{First-year Sloan Digital Sky Survey-II (SDSS-II)
  Supernova Results: Hubble Diagram and Cosmological Parameters},''
  \href{http://dx.doi.org/10.1088/0067-0049/185/1/32}{{\em Astrophys. J.
  Suppl.} {\bfseries 185} (2009) 32--84},
\href{http://arxiv.org/abs/0908.4274}{{arXiv:0908.4274 [astro-ph.CO]}}.

\bibitem{Vikhlinin:2008ym}
A.~Vikhlinin {\em et al.}, ``{Chandra Cluster Cosmology Project III:
  Cosmological Parameter Constraints},''
  \href{http://dx.doi.org/10.1088/0004-637X/692/2/1060}{{\em Astrophys. J.}
  {\bfseries 692} (2009) 1060--1074},
\href{http://arxiv.org/abs/0812.2720}{{arXiv:0812.2720 [astro-ph]}}.

\bibitem{Guth:1980zm}
A.~H. Guth, ``{The Inflationary Universe: A Possible Solution to the Horizon
  and Flatness Problems},''
\href{http://dx.doi.org/10.1103/PhysRevD.23.347}{{\em Phys. Rev.} {\bfseries
  D23} (1981) 347--356}.

\bibitem{Linde:1983gd}
A.~D. Linde, ``{Chaotic Inflation},''
\href{http://dx.doi.org/10.1016/0370-2693(83)90837-7}{{\em Phys. Lett.}
  {\bfseries B129} (1983) 177--181}.

\bibitem{Albrecht:1982wi}
A.~Albrecht and P.~J. Steinhardt, ``{Cosmology for Grand Unified Theories with
  Radiatively Induced Symmetry Breaking},''
\href{http://dx.doi.org/10.1103/PhysRevLett.48.1220}{{\em Phys. Rev. Lett.}
  {\bfseries 48} (1982) 1220--1223}.

\bibitem{Guth:1982ec}
A.~H. Guth and S.~Y. Pi, ``{Fluctuations in the New Inflationary Universe},''
\href{http://dx.doi.org/10.1103/PhysRevLett.49.1110}{{\em Phys. Rev. Lett.}
  {\bfseries 49} (1982) 1110--1113}.

\bibitem{Hawking:1982cz}
S.~W. Hawking, ``{The Development of Irregularities in a Single Bubble
  Inflationary Universe},''
\href{http://dx.doi.org/10.1016/0370-2693(82)90373-2}{{\em Phys. Lett.}
  {\bfseries B115} (1982) 295}.

\bibitem{Starobinsky:1982ee}
A.~A. Starobinsky, ``{Dynamics of Phase Transition in the New Inflationary
  Universe Scenario and Generation of Perturbations},''
\href{http://dx.doi.org/10.1016/0370-2693(82)90541-X}{{\em Phys. Lett.}
  {\bfseries B117} (1982) 175--178}.

\bibitem{Bardeen:1983qw}
J.~M. Bardeen, P.~J. Steinhardt, and M.~S. Turner, ``{Spontaneous Creation of
  Almost Scale - Free Density Perturbations in an Inflationary Universe},''
\href{http://dx.doi.org/10.1103/PhysRevD.28.679}{{\em Phys. Rev.} {\bfseries
  D28} (1983) 679}.

\bibitem{Salopek:1990jq}
D.~S. Salopek and J.~R. Bond, ``{Nonlinear evolution of long wavelength metric
  fluctuations in inflationary models},''
\href{http://dx.doi.org/10.1103/PhysRevD.42.3936}{{\em Phys. Rev.} {\bfseries
  D42} (1990) 3936--3962}.

\bibitem{Gangui:1993tt}
A.~Gangui, F.~Lucchin, S.~Matarrese, and S.~Mollerach, ``{The Three point
  correlation function of the cosmic microwave background in inflationary
  models},'' \href{http://dx.doi.org/10.1086/174421}{{\em Astrophys. J.}
  {\bfseries 430} (1994) 447--457},
\href{http://arxiv.org/abs/astro-ph/9312033}{{arXiv:astro-ph/9312033}}.

\bibitem{Komatsu:2001rj}
E.~Komatsu and D.~N. Spergel, ``{Acoustic signatures in the primary microwave
  background bispectrum},''
  \href{http://dx.doi.org/10.1103/PhysRevD.63.063002}{{\em Phys. Rev.}
  {\bfseries D63} (2001) 063002},
\href{http://arxiv.org/abs/astro-ph/0005036}{{arXiv:astro-ph/0005036}}.

\bibitem{Okamoto:2002ik}
T.~Okamoto and W.~Hu, ``{The Angular Trispectra of CMB Temperature and
  Polarization},'' \href{http://dx.doi.org/10.1103/PhysRevD.66.063008}{{\em
  Phys. Rev.} {\bfseries D66} (2002) 063008},
\href{http://arxiv.org/abs/astro-ph/0206155}{{arXiv:astro-ph/0206155}}.

\bibitem{Linde:1996gt}
A.~D. Linde and V.~F. Mukhanov, ``{Nongaussian isocurvature perturbations from
  inflation},'' \href{http://dx.doi.org/10.1103/PhysRevD.56.R535}{{\em Phys.
  Rev.} {\bfseries D56} (1997) 535--539},
\href{http://arxiv.org/abs/astro-ph/9610219}{{arXiv:astro-ph/9610219}}.

\bibitem{Lyth:2001nq}
D.~H. Lyth and D.~Wands, ``{Generating the curvature perturbation without an
  inflaton},'' \href{http://dx.doi.org/10.1016/S0370-2693(01)01366-1}{{\em
  Phys. Lett.} {\bfseries B524} (2002) 5--14},
\href{http://arxiv.org/abs/hep-ph/0110002}{{arXiv:hep-ph/0110002}}.

\bibitem{Lyth:2002my}
D.~H. Lyth, C.~Ungarelli, and D.~Wands, ``{The primordial density perturbation
  in the curvaton scenario},''
  \href{http://dx.doi.org/10.1103/PhysRevD.67.023503}{{\em Phys. Rev.}
  {\bfseries D67} (2003) 023503},
\href{http://arxiv.org/abs/astro-ph/0208055}{{arXiv:astro-ph/0208055}}.

\bibitem{Dvali:2003em}
G.~Dvali, A.~Gruzinov, and M.~Zaldarriaga, ``{A new mechanism for generating
  density perturbations from inflation},''
  \href{http://dx.doi.org/10.1103/PhysRevD.69.023505}{{\em Phys. Rev.}
  {\bfseries D69} (2004) 023505},
\href{http://arxiv.org/abs/astro-ph/0303591}{{arXiv:astro-ph/0303591}}.

\bibitem{Kofman:2003nx}
L.~Kofman, ``{Probing string theory with modulated cosmological
  fluctuations},''
\href{http://arxiv.org/abs/astro-ph/0303614}{{arXiv:astro-ph/0303614}}.

\bibitem{Buchbinder:2007at}
E.~I. Buchbinder, J.~Khoury, and B.~A. Ovrut, ``{Non-Gaussianities in New
  Ekpyrotic Cosmology},''
  \href{http://dx.doi.org/10.1103/PhysRevLett.100.171302}{{\em Phys. Rev.
  Lett.} {\bfseries 100} (2008) 171302},
\href{http://arxiv.org/abs/0710.5172}{{arXiv:0710.5172 [hep-th]}}.

\bibitem{Creminelli:2007aq}
P.~Creminelli and L.~Senatore, ``{A smooth bouncing cosmology with scale
  invariant spectrum},''
  \href{http://dx.doi.org/10.1088/1475-7516/2007/11/010}{{\em JCAP} {\bfseries
  0711} (2007) 010},
\href{http://arxiv.org/abs/hep-th/0702165}{{arXiv:hep-th/0702165}}.

\bibitem{Lehners:2007wc}
J.-L. Lehners and P.~J. Steinhardt, ``{Non-Gaussian Density Fluctuations from
  Entropically Generated Curvature Perturbations in Ekpyrotic Models},''
  \href{http://dx.doi.org/10.1103/PhysRevD.77.063533}{{\em Phys. Rev.}
  {\bfseries D77} (2008) 063533},
\href{http://arxiv.org/abs/0712.3779}{{arXiv:0712.3779 [hep-th]}}.

\bibitem{Maldacena:2002vr}
J.~M. Maldacena, ``{Non-Gaussian features of primordial fluctuations in single
  field inflationary models},'' {\em JHEP} {\bfseries 05} (2003) 013,
\href{http://arxiv.org/abs/astro-ph/0210603}{{arXiv:astro-ph/0210603}}.

\bibitem{Creminelli:2004yq}
P.~Creminelli and M.~Zaldarriaga, ``{Single field consistency relation for the
  3-point function},''
  \href{http://dx.doi.org/10.1088/1475-7516/2004/10/006}{{\em JCAP} {\bfseries
  0410} (2004) 006},
\href{http://arxiv.org/abs/astro-ph/0407059}{{arXiv:astro-ph/0407059}}.

\bibitem{Slosar:2008hx}
A.~Slosar, C.~Hirata, U.~Seljak, S.~Ho, and N.~Padmanabhan, ``{Constraints on
  local primordial non-Gaussianity from large scale structure},''
  \href{http://dx.doi.org/10.1088/1475-7516/2008/08/031}{{\em JCAP} {\bfseries
  0808} (2008) 031},
\href{http://arxiv.org/abs/0805.3580}{{arXiv:0805.3580 [astro-ph]}}.

\bibitem{Fergusson:2010gn}
J.~R. Fergusson, D.~M. Regan, and E.~P.~S. Shellard, ``{Optimal Trispectrum
  Estimators and WMAP Constraints},''
\href{http://arxiv.org/abs/1012.6039}{{arXiv:1012.6039 [astro-ph.CO]}}.

\bibitem{Desjacques:2008vf}
V.~Desjacques, U.~Seljak, and I.~Iliev, ``{Scale-dependent bias induced by
  local non-Gaussianity: A comparison to N-body simulations},''
\href{http://arxiv.org/abs/0811.2748}{{arXiv:0811.2748 [astro-ph]}}.

\bibitem{Sasaki:2006kq}
M.~Sasaki, J.~Valiviita, and D.~Wands, ``{Non-gaussianity of the primordial
  perturbation in the curvaton model},''
  \href{http://dx.doi.org/10.1103/PhysRevD.74.103003}{{\em Phys. Rev.}
  {\bfseries D74} (2006) 103003},
\href{http://arxiv.org/abs/astro-ph/0607627}{{arXiv:astro-ph/0607627}}.

\bibitem{Ichikawa:2008iq}
K.~Ichikawa, T.~Suyama, T.~Takahashi, and M.~Yamaguchi, ``{Non-Gaussianity,
  Spectral Index and Tensor Modes in Mixed Inflaton and Curvaton Models},''
  \href{http://dx.doi.org/10.1103/PhysRevD.78.023513}{{\em Phys. Rev.}
  {\bfseries D78} (2008) 023513},
\href{http://arxiv.org/abs/0802.4138}{{arXiv:0802.4138 [astro-ph]}}.

\bibitem{Enqvist:2008gk}
K.~Enqvist and T.~Takahashi, ``{Signatures of Non-Gaussianity in the Curvaton
  Model},'' \href{http://dx.doi.org/10.1088/1475-7516/2008/09/012}{{\em JCAP}
  {\bfseries 0809} (2008) 012},
\href{http://arxiv.org/abs/0807.3069}{{arXiv:0807.3069 [astro-ph]}}.

\bibitem{Huang:2008zj}
Q.-G. Huang, ``{Curvaton with Polynomial Potential},''
  \href{http://dx.doi.org/10.1088/1475-7516/2008/11/005}{{\em JCAP} {\bfseries
  0811} (2008) 005},
\href{http://arxiv.org/abs/0808.1793}{{arXiv:0808.1793 [hep-th]}}.

\bibitem{Chingangbam:2009xi}
P.~Chingangbam and Q.-G. Huang, ``{The Curvature Perturbation in the Axion-type
  Curvaton Model},''
  \href{http://dx.doi.org/10.1088/1475-7516/2009/04/031}{{\em JCAP} {\bfseries
  0904} (2009) 031},
\href{http://arxiv.org/abs/0902.2619}{{arXiv:0902.2619 [astro-ph.CO]}}.

\bibitem{Huang:2009vk}
Q.-G. Huang, ``{A geometric description of the non-Gaussianity generated at the
  end of multi-field inflation},''
  \href{http://dx.doi.org/10.1088/1475-7516/2009/06/035}{{\em JCAP} {\bfseries
  0906} (2009) 035},
\href{http://arxiv.org/abs/0904.2649}{{arXiv:0904.2649 [hep-th]}}.

\bibitem{Byrnes:2009qy}
C.~T. Byrnes and G.~Tasinato, ``{Non-Gaussianity beyond slow roll in
  multi-field inflation},''
  \href{http://dx.doi.org/10.1088/1475-7516/2009/08/016}{{\em JCAP} {\bfseries
  0908} (2009) 016},
\href{http://arxiv.org/abs/0906.0767}{{arXiv:0906.0767 [astro-ph.CO]}}.

\bibitem{Dalal:2007cu}
N.~Dalal, O.~Dore, D.~Huterer, and A.~Shirokov, ``{The imprints of primordial
  non-gaussianities on large- scale structure: scale dependent bias and
  abundance of virialized objects},''
  \href{http://dx.doi.org/10.1103/PhysRevD.77.123514}{{\em Phys. Rev.}
  {\bfseries D77} (2008) 123514},
\href{http://arxiv.org/abs/0710.4560}{{arXiv:0710.4560 [astro-ph]}}.

\bibitem{Xia:2011hj}
J.-Q. Xia, C.~Baccigalupi, S.~Matarrese, L.~Verde, and M.~Viel, ``{Constraints
  on Primordial Non-Gaussianity from Large Scale Structure Probes},''
\href{http://arxiv.org/abs/1104.5015}{{arXiv:1104.5015 [astro-ph.CO]}}.

\bibitem{Cunha:2010zz}
C.~Cunha, D.~Huterer, and O.~Dore, ``{Primordial non-Gaussianity from the
  covariance of galaxy cluster counts},''
  \href{http://dx.doi.org/10.1103/PhysRevD.82.023004}{{\em Phys. Rev.}
  {\bfseries D82} (2010) 023004},
\href{http://arxiv.org/abs/1003.2416}{{arXiv:1003.2416 [astro-ph.CO]}}.

\bibitem{Hamaus:2011dq}
N.~Hamaus, U.~Seljak, and V.~Desjacques, ``{Optimal Constraints on Local
  Primordial Non-Gaussianity from the Two-Point Statistics of Large-Scale
  Structure},''
\href{http://arxiv.org/abs/1104.2321}{{arXiv:1104.2321 [astro-ph.CO]}}.

\bibitem{Matarrese:2008nc}
S.~Matarrese and L.~Verde, ``{The effect of primordial non-Gaussianity on halo
  bias},'' \href{http://dx.doi.org/10.1086/587840}{{\em Astrophys. J.}
  {\bfseries 677} (2008) L77},
\href{http://arxiv.org/abs/0801.4826}{{arXiv:0801.4826 [astro-ph]}}.

\bibitem{Giannantonio:2009ak}
T.~Giannantonio and C.~Porciani, ``{Structure formation from non-Gaussian
  initial conditions: multivariate biasing, statistics, and comparison with N-
  body simulations},'' \href{http://dx.doi.org/10.1103/PhysRevD.81.063530}{{\em
  Phys. Rev.} {\bfseries D81} (2010) 063530},
\href{http://arxiv.org/abs/0911.0017}{{arXiv:0911.0017 [astro-ph.CO]}}.

\bibitem{Grossi:2007ry}
M.~Grossi, K.~Dolag, E.~Branchini, S.~Matarrese, and L.~Moscardini,
  ``{Evolution of Massive Haloes in non-Gaussian Scenarios},'' {\em Mon. Not.
  Roy. Astron. Soc.} {\bfseries 382} (2007) 1261,
\href{http://arxiv.org/abs/0707.2516}{{arXiv:0707.2516 [astro-ph]}}.

\bibitem{Pillepich:2008ka}
A.~Pillepich, C.~Porciani, and O.~Hahn, ``{Universal halo mass function and
  scale-dependent bias from N-body simulations with non-Gaussian initial
  conditions},''
\href{http://arxiv.org/abs/0811.4176}{{arXiv:0811.4176 [astro-ph]}}.

\bibitem{Desjacques:2011mq}
V.~Desjacques, D.~Jeong, and F.~Schmidt, ``{Non-Gaussian Halo Bias Re-examined:
  Mass-dependent Amplitude from the Peak-Background Split and Thresholding},''
\href{http://arxiv.org/abs/1105.3628}{{arXiv:1105.3628 [astro-ph.CO]}}.

\bibitem{Dunkley:2008ie}
{ WMAP} Collaboration, J.~Dunkley {\em et al.}, ``{Five-Year Wilkinson
  Microwave Anisotropy Probe (WMAP) Observations: Likelihoods and Parameters
  from the WMAP data},''
  \href{http://dx.doi.org/10.1088/0067-0049/180/2/306}{{\em Astrophys. J.
  Suppl.} {\bfseries 180} (2009) 306--329},
\href{http://arxiv.org/abs/0803.0586}{{arXiv:0803.0586 [astro-ph]}}.

\bibitem{Lewis:1999bs}
A.~Lewis, A.~Challinor, and A.~Lasenby, ``{Efficient Computation of CMB
  anisotropies in closed FRW models},''
  \href{http://dx.doi.org/10.1086/309179}{{\em Astrophys. J.} {\bfseries 538}
  (2000) 473--476},
\href{http://arxiv.org/abs/astro-ph/9911177}{{arXiv:astro-ph/9911177}}.

\bibitem{LoVerde:2011iz}
M.~LoVerde and K.~M. Smith, ``{The Non-Gaussian Halo Mass Function with
  $f_{NL}$, $g_{NL}$ and $\tau_{NL}$},''
\href{http://arxiv.org/abs/1102.1439}{{arXiv:1102.1439 [astro-ph.CO]}}.

\bibitem{simone_stochasticity}
S.~Ferraro, M.~LoVerde, and K.~M. Smith, ``{Large-scale halo stochasticity and
  primordial non-Gaussianity},'' \href{http://arxiv.org/abs/to appear}{{to
  appear}}.

\bibitem{Matarrese:2000iz}
S.~Matarrese, L.~Verde, and R.~Jimenez, ``{The abundance of high-redshift
  objects as a probe of non- Gaussian initial conditions},''
  \href{http://dx.doi.org/10.1086/309412}{{\em Astrophys. J.} {\bfseries 541}
  (2000) 10},
\href{http://arxiv.org/abs/astro-ph/0001366}{{arXiv:astro-ph/0001366}}.

\bibitem{LoVerde:2007ri}
M.~LoVerde, A.~Miller, S.~Shandera, and L.~Verde, ``{Effects of Scale-Dependent
  Non-Gaussianity on Cosmological Structures},''
  \href{http://dx.doi.org/10.1088/1475-7516/2008/04/014}{{\em JCAP} {\bfseries
  0804} (2008) 014},
\href{http://arxiv.org/abs/0711.4126}{{arXiv:0711.4126 [astro-ph]}}.

\bibitem{Chongchitnan:2010xz}
S.~Chongchitnan and J.~Silk, ``{A Study of High-Order Non-Gaussianity with
  Applications to Massive Clusters and Large Voids},''
  \href{http://dx.doi.org/10.1088/0004-637X/724/1/285}{{\em Astrophys.J.}
  {\bfseries 724} (2010) 285--295},
  \href{http://arxiv.org/abs/1007.1230}{{arXiv:1007.1230 [astro-ph.CO]}}.

\bibitem{Lam:2009nb}
T.~Y. Lam and R.~K. Sheth, ``{Halo abundances in the $f_{nl}$ model},''
\href{http://arxiv.org/abs/0905.1702}{{arXiv:0905.1702 [astro-ph.CO]}}.

\bibitem{Maggiore:2009rx}
M.~Maggiore and A.~Riotto, ``{The halo mass function from the excursion set
  method. III. First principle derivation for non-Gaussian theories},''
\href{http://arxiv.org/abs/0903.1251}{{arXiv:0903.1251 [astro-ph.CO]}}.

\bibitem{DeSimone:2010mu}
A.~De~Simone, M.~Maggiore, and A.~Riotto, ``{Excursion Set Theory for generic
  moving barriers and non- Gaussian initial conditions},''
\href{http://arxiv.org/abs/1007.1903}{{arXiv:1007.1903 [astro-ph.CO]}}.

\bibitem{D'Amico:2010ta}
G.~D'Amico, M.~Musso, J.~Norena, and A.~Paranjape, ``{An Improved Calculation
  of the Non-Gaussian Halo Mass Function},''
\href{http://arxiv.org/abs/1005.1203}{{arXiv:1005.1203 [astro-ph.CO]}}.

\bibitem{Press:1973iz}
W.~H. Press and P.~Schechter, ``{Formation of galaxies and clusters of galaxies
  by selfsimilar gravitational condensation},''
\href{http://dx.doi.org/10.1086/152650}{{\em Astrophys. J.} {\bfseries 187}
  (1974) 425--438}.

\bibitem{Sheth:1999mn}
R.~K. Sheth and G.~Tormen, ``{Large scale bias and the peak background
  split},'' \href{http://dx.doi.org/10.1046/j.1365-8711.1999.02692.x}{{\em Mon.
  Not. Roy. Astron. Soc.} {\bfseries 308} (1999) 119},
\href{http://arxiv.org/abs/astro-ph/9901122}{{arXiv:astro-ph/9901122}}.

\bibitem{Warren:2005ey}
M.~S. Warren, K.~Abazajian, D.~E. Holz, and L.~Teodoro, ``{Precision
  Determination of the Mass Function of Dark Matter Halos},''
  \href{http://dx.doi.org/10.1086/504962}{{\em Astrophys. J.} {\bfseries 646}
  (2006) 881--885},
\href{http://arxiv.org/abs/astro-ph/0506395}{{arXiv:astro-ph/0506395}}.

\bibitem{Cole:1989vx}
S.~Cole and N.~Kaiser, ``{Biased clustering in the cold dark matter
  cosmogony},''
{\em Mon. Not. Roy. Astron. Soc.} {\bfseries 237} (1989) 1127--1146.

\bibitem{Desjacques:2009jb}
V.~Desjacques and U.~Seljak, ``{Signature of primordial non-Gaussianity of
  $\phi^3$-type in the mass function and bias of dark matter haloes},''
  \href{http://dx.doi.org/10.1103/PhysRevD.81.023006}{{\em Phys. Rev.}
  {\bfseries D81} (2010) 023006},
\href{http://arxiv.org/abs/0907.2257}{{arXiv:0907.2257 [astro-ph.CO]}}.

\bibitem{Matarrese:1986et}
S.~Matarrese, F.~Lucchin, and S.~A. Bonometto, ``{A Path Integral Approach to
  Large Scale Matter Distribution Originated by Non-Gaussian Fluctuations},''
{\em Astrophys. J.} {\bfseries 310} (1986) L21--l26.

\bibitem{Springel:2005mi}
V.~Springel, ``{The cosmological simulation code GADGET-2},''
  \href{http://dx.doi.org/10.1111/j.1365-2966.2005.09655.x}{{\em Mon. Not. Roy.
  Astron. Soc.} {\bfseries 364} (2005) 1105--1134},
\href{http://arxiv.org/abs/astro-ph/0505010}{{arXiv:astro-ph/0505010}}.

\bibitem{Zeldovich:1969sb}
Y.~B. Zeldovich, ``{Gravitational instability: An Approximate theory for large
  density perturbations},''
{\em Astron. Astrophys.} {\bfseries 5} (1970) 84--89.

\bibitem{Crocce:2006ve}
M.~Crocce, S.~Pueblas, and R.~Scoccimarro, ``{Transients from Initial
  Conditions in Cosmological Simulations},''
  \href{http://dx.doi.org/10.1111/j.1365-2966.2006.11040.x}{{\em Mon. Not. Roy.
  Astron. Soc.} {\bfseries 373} (2006) 369--381},
\href{http://arxiv.org/abs/astro-ph/0606505}{{arXiv:astro-ph/0606505}}.

\bibitem{Frenk:1988zz}
C.~S. Frenk, S.~D.~M. White, M.~Davis, and G.~Efstathiou, ``{The formation of
  dark halos in a universe dominated by cold dark matter},''
{\em Astrophys. J.} {\bfseries 327} (1988) 507--525.

\bibitem{Smith:2010gx}
K.~M. Smith and M.~LoVerde, ``{Local stochastic non-Gaussianity and N-body
  simulations},''
\href{http://arxiv.org/abs/1010.0055}{{arXiv:1010.0055 [astro-ph.CO]}}.

\bibitem{Reid:2010vc}
B.~A. Reid, L.~Verde, K.~Dolag, S.~Matarrese, and L.~Moscardini,
  ``{Non-Gaussian halo assembly bias},''
  \href{http://dx.doi.org/10.1088/1475-7516/2010/07/013}{{\em JCAP} {\bfseries
  1007} (2010) 013},
\href{http://arxiv.org/abs/1004.1637}{{arXiv:1004.1637 [astro-ph.CO]}}.

\end{thebibliography}\endgroup

\appendix

\section{Barrier model calculations}
\label{app:barrier}

In this appendix, we give details of the calculation of the halo mass function and large-scale bias
(Eqs.~(\ref{eq:barrier_mf})--(\ref{eq:barrier_b2})) in the barrier crossing model, to first order in $\fnl$, $\gnl$.

First, consider evaluation of the integrals in Eqs.~(\ref{eq:p0_def}),~(\ref{eq:xi0_def}).
Primordial non-Gaussianity enters the calculation by perturbing the PDFs which appear from Gaussian distributions.
This perturbation can be written down explicitly using the Edgeworth expansion, which represents
the PDF as a power series in cumulants.
The Edgeworth expansion for the 1-variable PDF $p(\delta_M')$ is:
\ba
p(\delta_M') &=& \exp \left( \sum_{n\ge 3} \frac{(-1)^n}{n!} \kappa_n(M) \sigma_M^n \frac{\partial^n}{\partial\delta_M^{'n}} \right)
    \frac{1}{(2\pi)^{1/2}\sigma_M} e^{-\delta_M^{'2}/(2\sigma_M^2)} \nn \\
 &=& \frac{1}{(2\pi)^{1/2}\sigma_M} e^{-\delta_M^{'2}/(2\sigma_M^2)} 
      \left( 1 + \frac{\kappa_3(M)}{6} H_3(\nu) + \frac{\kappa_4(M)}{24} H_4(\nu) + \cdots \right) \nn \\
 &=& \frac{1}{(2\pi)^{1/2}\sigma_M} e^{-\delta_M^{'2}/(2\sigma_M^2)} 
      \left( 1 + \fnl \frac{\kappa_3^{(1)}(M)}{6} H_3(\nu) + \gnl \frac{\kappa_4^{(1)}(M)}{24} H_4(\nu) + \cdots \right)
\ea
where we have kept terms of first order in $\fnl$, $\gnl$.
We can now compute $p_0$ by plugging into the definition~(\ref{eq:p0_def}):
\be
p_0 = \frac{1}{2} \erfc\left( \frac{\nu}{\sqrt{2}} \right) 
   + \fnl \frac{\kappa_3^{(1)}(M)}{6} \frac{e^{-\nu^2/2}}{(2\pi)^{1/2}} H_2(\nu)
   + \gnl \frac{\kappa_4^{(1)}(M)}{24} \frac{e^{-\nu^2/2}}{(2\pi)^{1/2}} H_3(\nu)  \label{eq:barrier_p0}
\ee
Armed with this expression, it is easy to compute $n(M) = -2\rho_m/M (dp_0/dM)$, obtaining the
form of the mass function in Eq.~(\ref{eq:barrier_mf}).

Moving on to the 2-variable PDF $p(\delta_{\rm lin}, \delta_M')$, the Edgeworth expansion is:
\ba
p(\delta_{\rm lin}, \delta_M')
  &=& \exp\left( \sigma_{\rm lin} \sigma_M \kappa_{1,1} \frac{\partial^2}{\partial\delta_{\rm lin} \, \partial\delta_M'} 
         + \sum_{\substack{m,n \\ m+n\ge 3}} \frac{(-1)^{m+n}}{m!n!} \sigma_{\rm lin}^m \sigma_M^n \kappa_{m,n}
              \frac{\partial^{m+n}}{\partial\delta_{\rm lin}^m \partial\delta_M^{'n}} \right) \nn \\
    && \hspace{1.5cm} \times  \frac{1}{2\pi \sigma_{\rm lin} \sigma_M}
        \exp\left( -\frac{\delta_{\rm lin}^2}{2\sigma_{\rm lin}^2} - \frac{\delta_M^{'2}}{2\sigma_M^2} \right)  \label{eq:edgeworth2}
\ea
where $\sigma_{\rm lin} = \langle \delta_{\rm lin}^2 \rangle^{1/2}$ and the cumulant $\kappa_{m,n}$ is defined by:\footnote{A technical
point: $\sigma_{\rm lin}$ is formally infinite, but it will cancel from the final results in Eqs.~(\ref{eq:barrier_b0})--(\ref{eq:barrier_b2}).
One could make $\sigma_{\rm lin}$ finite by introducing a smoothing scale $R$ for the matter field, and take the limit $R\rightarrow 0$
at the end of the calculation.}
\be
\kappa_{m,n}(M,r) = \frac{\left\langle (\delta_{\rm lin})^m (\delta_M')^n \right\rangle_{\rm conn}}{\sigma_{\rm lin}^m \sigma_M^n}
\ee
Note that the cumulant $\kappa_n(M)$ defined previously in Eq.~(\ref{eq:kappa_def}) is equal to $\kappa_{0,n}(M,r)$.

Keeping the first few terms in the Edgeworth expansion:\footnote{The choice of terms to keep was dictated
by the following considerations.  Only terms with precisely one $\delta_{\rm lin}$ derivative will give nonzero contributions
to the integral $\int_{-\infty}^\infty d\delta_{\rm lin}\, \delta_{\rm lin} p(\delta_{\rm lin}, \delta_M')$ appearing in $\xi_{mh}(r)$,
so we have only kept these terms.  (Terms with two or more derivatives would contribute to the halo-halo correlation function $\xi_{hh}(r)$, so
they may be relevant for halo stochasticity.)  We have also omitted terms whose leading contribution is second-order or
higher in $\fnl$ and $\gnl$.}
\ba
p(\delta_{\rm lin}, \delta_M')
  &=& \frac{1}{2\pi \sigma_{\rm lin} \sigma_M} \exp\left( -\frac{\delta_{\rm lin}^2}{2\sigma_{\rm lin}^2} - \frac{\delta_M^{'2}}{2\sigma_M^2} \right) \nn  \\
   && \hspace{0.5cm} \times 
         \Bigg( 1 + \frac{\kappa_{1,1}(M,r)}{\sigma_{\rm lin}} \delta_{\rm lin} \!\left( \frac{\delta_M'}{\sigma_M} \right)
               + \frac{\kappa_{1,2}(M,r)}{2\sigma_{\rm lin}} \delta_{\rm lin} H_2\!\left( \frac{\delta_M'}{\sigma_M} \right) \nn \\
   && \hspace{1cm}
               + \frac{\kappa_{1,3}(M,r)}{6\sigma_{\rm lin}} \delta_{\rm lin} H_3\!\left( \frac{\delta_M'}{\sigma_M} \right) 
               + \frac{\kappa_{1,1}(M,r) \kappa_{0,3}(M)}{6 \sigma_{\rm lin}} \delta_{\rm lin} H_4\!\left( \frac{\delta_M'}{\sigma_M} \right) \nn \\
   && \hspace{1cm}
               + \frac{\kappa_{1,1}(M,r) \kappa_{0,4}(M)}{24 \sigma_{\rm lin}} \delta_{\rm lin} H_5\!\left( \frac{\delta_M'}{\sigma_M} \right)
               + \cdots \Bigg)
\ea
we compute $\xi_0(r)$ by integrating Eq.~(\ref{eq:xi0_def}) term by term, obtaining:
\ba
\xi_0(r) &=& \frac{\sigma_{\rm lin} e^{-\nu^2/2}}{(2\pi)^{1/2}} \Bigg(
   \kappa_{1,1}(M,r) + \frac{\kappa_{1,2}(M,r)}{2} \nu + \frac{\kappa_{1,3}(M,r)}{6} H_2(\nu) \nn \\
&& \hspace{1cm} + \frac{\kappa_{1,1}(M,r) \kappa_3(M)}{6} H_3(\nu) 
 + \frac{\kappa_{1,1}(M,r) \kappa_4(M)}{24} H_4(\nu) \Bigg)
\ea
To make further progress, we convert the correlation function to a power spectrum $P_0(k) = \int d^3\r\, e^{i\k\cdot\r} \xi_0(r)$,
and keep only the leading behavior of each term in the long-wavelength limit $k\rightarrow 0$.
\ba
\int d^3\r\, e^{i\k\cdot\r} \kappa_{1,2}(M,r)
  &=& \frac{1}{\sigma_{\rm lin} \sigma_M^2} \int \frac{d^3\q\,d^3\q'}{(2\pi)^6} W_M(q) W_M(q')\, \big\langle \delta(\k) \delta(\q) \delta(-\q') \big\rangle \nn \\
  & \rightarrow & \frac{4 \fnl}{\sigma_{\rm lin}} \frac{P(k)}{\alpha(k)}
\ea
\ba
\int d^3\r\, e^{i\k\cdot\r} \kappa_{1,3}(M,r)
  &=& \frac{1}{\sigma_{\rm lin} \sigma_M^3} \int \frac{d^3\q\,d^3\q'\,d^3\q''}{(2\pi)^9} W_M(q) W_M(q') W_M(q'') \nn \\
   && \hspace{1cm} \times \big\langle \delta(\k) \delta(\q) \delta(\q') \delta(-\q'') \big\rangle_{\rm conn} \nn \\
  & \rightarrow & \frac{18 \gnl P(k)}{\sigma_{\rm lin} \sigma_M^3 \alpha(k)} 
      \int \frac{d^3\q\,d^3\q'}{(2\pi)^6} W_M(q) W_M(q') W_M(|\q+\q'|) \nn \\
  && \hspace{1cm} \times \frac{P(q) P(q') \alpha(|\q+\q'|)}{\alpha(q) \alpha(q')} \nn \\
  & = & \frac{3 \gnl}{\sigma_{\rm lin}} \kappa_3^{(1)}(M) \left( \frac{P(k)}{\alpha(k)} \right)
\ea
where ``$\rightarrow$'' denotes the $k\rightarrow 0$ limit, and we have used Eq.~(\ref{eq:kappa3_explicit}) to simplify the last line.
Putting this together, we find the following expression for $P_0(k)$ in the $k\rightarrow 0$ limit:
\ba
P_0(k) &=& \frac{e^{-\nu^2/2}}{(2\pi)^{1/2}} \Bigg[ \frac{P(k)}{\sigma_M} \left(1 + \fnl \frac{\kappa_3^{(1)}(M)}{6} H_3(\nu)
  + \gnl \frac{\kappa_4^{(1)}(M)}{24} H_4(\nu) \right) \nn \\
  && \hspace{3cm} + 2\nu \fnl \frac{P(k)}{\alpha(k)} + \kappa_3^{(1)}(M) \frac{H_2(\nu)}{2} \gnl \frac{P(k)}{\alpha(k)} \Bigg]  \label{eq:barrier_P0}
\ea
The halo bias in a narrow mass range is given by the derivative:
\be
b(k) = \frac{dP_0(k)/dM}{(dp_0/dM) P(k)} + 1
\ee
where the ``+1'' converts Lagrangian to Eulerian bias.
Plugging in the forms of $p_0$, $P_0$ in Eqs.~(\ref{eq:barrier_p0}),~(\ref{eq:barrier_P0}), a long but straightforward calculation now
gives the halo bias in the form given in the text (Eqs.~(\ref{eq:barrier_bias})--(\ref{eq:barrier_b2})).

\end{document}